\documentclass[letterpaper]{article} 
\usepackage{aaai23}  
\usepackage{times}  
\usepackage{helvet}  
\usepackage{courier}  
\usepackage[hyphens]{url}  
\usepackage{graphicx} 
\usepackage{natbib}  
\usepackage{caption} 
\newcommand{\M}{SCI}
\newcommand{\tabfontsize}{\fontsize{8.2pt}{\baselineskip}\selectfont}

\usepackage{amsmath,amssymb}
\usepackage{subfigure}
\usepackage{color}
\usepackage{hyperref}
\usepackage{float}
\title{\M: A Spectrum Concentrated Implicit Neural Compression for Biomedical Data}

\author{
    Runzhao Yang\textsuperscript{\rm 1},
    Tingxiong Xiao\textsuperscript{\rm 1},
    Yuxiao Cheng\textsuperscript{\rm 1},
    Qianni Cao\textsuperscript{\rm 2},
    Jinyuan Qu\textsuperscript{\rm 1},
    Jinli Suo\textsuperscript{\rm 1}\textsuperscript{\rm 3},
    Qionghai Dai\textsuperscript{\rm 1}
}
\affiliations{
    \textsuperscript{\rm 1}Department of Automation, Tsinghua University, Beijing 100084, China\\

    \textsuperscript{\rm 2}Department of Electrical Engineering, Tsinghua University, Beijing 100084, China\\
    \textsuperscript{\rm 3}jlsuo@tsinghua.edu.cn
}

\usepackage{bibentry}

\begin{document}

\maketitle

\begin{abstract}
Massive collection and explosive growth of biomedical data, demands effective compression for efficient storage, transmission and sharing.
Readily available visual data compression techniques have been studied extensively but tailored for natural images/videos, and thus show limited performance on biomedical data which are of different features and larger diversity. 
Emerging implicit neural representation (INR) is gaining momentum and demonstrates high promise for fitting diverse visual data in target-data-specific manner, but a general compression scheme covering diverse biomedical data is so far absent. 
To address this issue, we firstly derive a mathematical explanation for INR's spectrum concentration property and an analytical insight on the design of INR based compressor. 
Further, we propose a Spectrum Concentrated Implicit neural compression (\M) which adaptively partitions the complex biomedical data into blocks matching INR's concentrated spectrum envelop, and design a funnel shaped neural network capable of representing each block with a small number of parameters. 
Based on this design, we conduct compression via optimization under given budget and allocate the available parameters with high representation accuracy. 
The experiments show \M's superior performance to state-of-the-art methods including commercial compressors, data-driven ones, and INR based counterparts on diverse biomedical data.
The source code can be found at \href{https://github.com/RichealYoung/ImplicitNeuralCompression.git}{https://github.com/RichealYoung/ImplicitNeuralCompression.git}.

\end{abstract}

\section{Introduction}
\begin{figure}[t]
\centering
\vspace{-2mm}
\includegraphics[width=\linewidth]{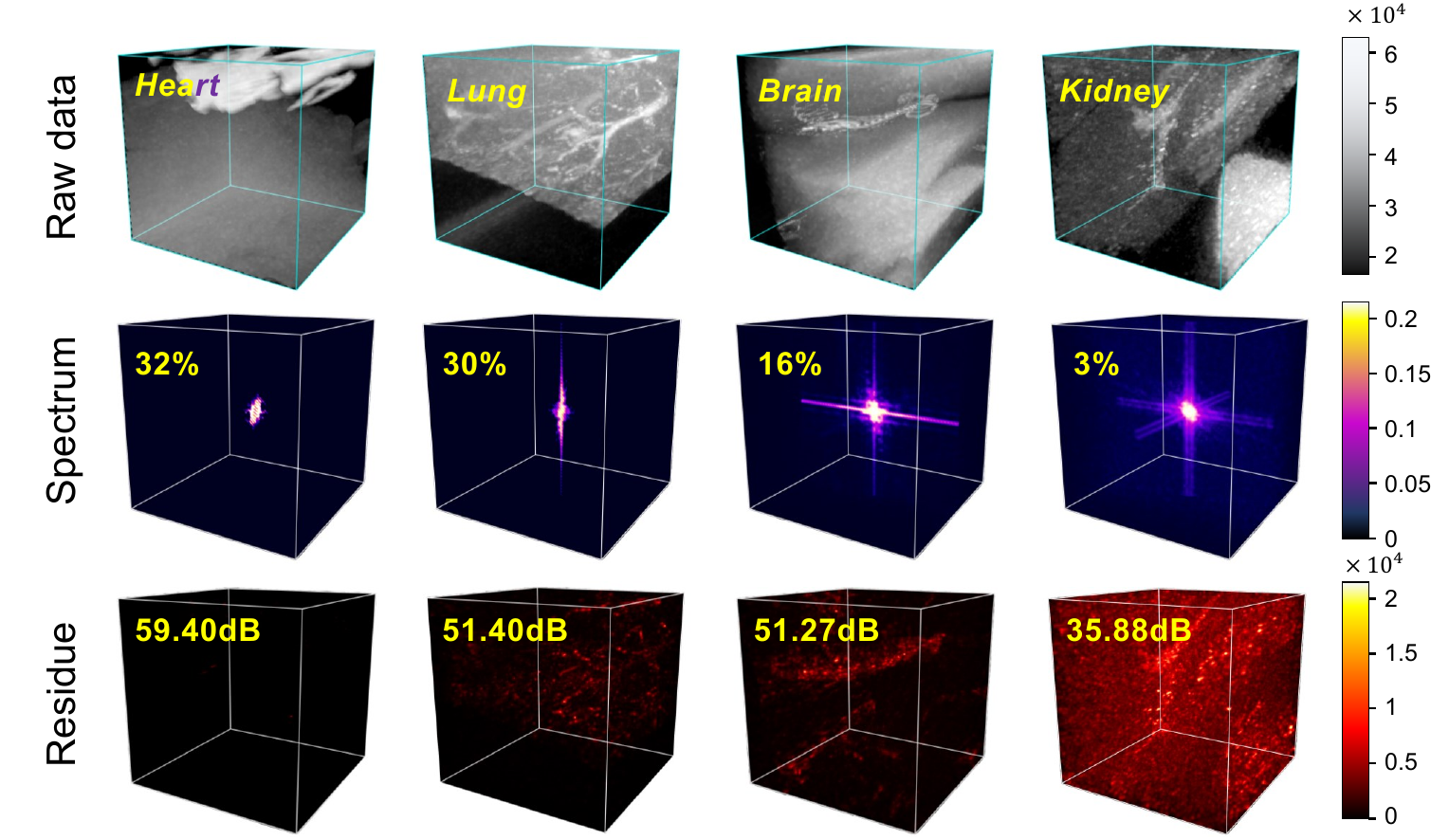}
\caption{INR's diverse compression capability on different biomedical data (from HiP-CT dataset), under $64\times$ compression ratio.
Top: several representative CT data of different organs, which are of increasing complexities from left to right. Middle: the corresponding spectrum (Fourier coefficients), with the percentage of non-zero entries labeled. Bottom: the residue after compression, which increases dramatically with the spreading range of the spectrum, with the reconstruction fidelity labeled in terms of PSNR.}
\vspace{-5mm}
\label{fig:motivation}
\end{figure}

Over decades, digital visual information is  most commonly represented as discrete intensity values over a regularly spaced grid. Such straightforward representation is highly redundant, so a bunch of compression tools have been developed, such as MPEG/JPEG \cite{wallace1992jpeg}, H.264 \cite{wiegand2003overview}, HEVC \cite{sullivan2012overview}, etc. In spite of their successful applications and widespread acceptance, most of these compression techniques are tailored for natural  images/videos and of limited performance in many domain specific visual data, such as biomedical images/volumes, which are of different characteristics and larger diversity than natural  scenes.

With the resurgence of deep neural networks, a type of new representation of visual data is emerging---Implicit Neural Representation (INR), which uses a neural network to parameterize the data with an implicit continuous function. Specifically, for a visual data, the intensity at a voxel coordinate can be described by a neural function. Different from conventional discrete grid-based representations, INR is defined over a continuous space, and thus better matches the continuity property of the target visual information and is intrinsically independent of the discretization levels. Besides, according to the universal approximation theorem of neural networks, an INR implemented with Multi-Layer Perceptron (MLP) can fit arbitrarily complex functions given sufficient parameters \cite{gallant1988there,hornik1989multilayer,hornik1991approximation,cybenko1989approximation,leshno1993multilayer,pinkus1999approximation}, which promises high fidelity representation \cite{hornik1989multilayer}. 
Under the INR representation, if we can model the underlying correlations among the neural functions of a target visual data at a set of positions, a compact INR representation can be achieved \cite{lu2021compressive,chen2021nerv}. Such INR based compression is obtaining better performance than conventional techniques on natural  images/videos, and holds big promise for biomedical data (or other special domains) via pursuing optimal data specific INR under parameter budget.

In spite of the high performance on natural visual data, INR based compression does not perform equally well on complex biomedical data like CT, which is of diverse spectrum distribution. As shown in Figure~\ref{fig:motivation}, the compression capability significantly decreases as the spectrum spreads, which leaves following key issues open for exploration:
\begin{itemize}
\vspace{-1mm}
    \item Why and how does a broad spectrum degrade INR's compression capability?
    \item How to keep INR's high compression capability on complex data with a broad spectrum?
\end{itemize}

To address the above issues, we start from analyzing the representation fidelity of INR from the spectrum perspective. The analysis reveals that INR is ``spectrum concentrated" and thus of low accuracy for data with spectrum beyond INR's spectrum envelop, which explicitly answers the first question well.
Following this, we propose spectrum concentrated implicit neural compression (\M) to boost the compression capability of INR in three ways. Firstly,  we develop an adaptive partitioning strategy to divide the target data with wide spreading spectrum into spectrum concentrated blocks with similar spectrum coverage. Secondly, we design a funnel shaped INR structure for efficient data compression, in which the number of neurons in each layer is elaborately designed for covering the spectrum of the target data block. Thirdly we establish an allocation strategy for the available number of INR parameters among the blocks.

Experimentally, we evaluated \M's performance and validated its wide applicability on biological and medical data. For biological data we used the brain-wide neurons and vessels of mouse, which have sparsely dispersed structures and their brightness distributions vary greatly in the observed data, especially that certain fine structures have faint brightness. For medical data we used HiP-CT \cite{walsh2021imaging}, a public dataset containing data volumes of human organs from the tissue to cellular scales, which are of large diversity in both appearance (e.g., structure, contrast, brightness, and noise levels). All these biomedical data are of broad spectrum range.
The results show that \M~can significantly improve the compression capability of the INR based compression (NeRF \cite{nerf}, SIREN \cite{siren}, and NeRV \cite{chen2021nerv}), and \M~even outperforms commercial compression techniques (JPEG \cite{wallace1992jpeg}, H.264 \cite{wiegand2003overview}, and HEVC \cite{sullivan2012overview}), and data-driven deep learning approaches (DVC \cite{lu2019dvc}), SGA+BB \cite{yang20improving}, and SSF~\cite{agustsson2020scale}), especially on the complex data with broad spectrum.
\vspace{2mm}

To summarize, the technical contributions are as follows:
\begin{itemize}
\vspace{-1mm}
    \item A mathematical explanation to INR's spectrum concentration property and theoretical analysis on how broad spectrum degrades INR based compression quality.
    \item An adaptive partitioning strategy, which divides data with spreading spectrum into blocks fit in the INR's spectrum envelop.
    \item An elaborately designed INR architecture capable of representing the blocks with small number of parameters.
    \item  A parameter allocation strategy adapting to the spectral coverage of data blocks to achieve best compression.
    \item \M~outperforms existing INR based compressions, data-driven deep learning  approaches and even widely used commercial compression techniques on biomedical data.
\end{itemize}

\section{Related work}
\noindent\textbf{Data compression.}\quad Data compression is fundamental for data sharing and storage. The community has witnessed the great progress in this field, from the handcrafted transform coding based compression methods such as JPEG \cite{wallace1992jpeg} and JPEG2000 \cite{skodras2001jpeg}, to the predictive coding compression methods, like H.264 \cite{wiegand2003overview} and HEVC \cite{sullivan2012overview}. These techniques have been widely and successfully used in compression for natural scenes. 
Built on conventional coding methodology, researchers attempt to use deep learning-based methods \cite{lu2019dvc,minnen18joint,johannes18variational,yang20improving}, which take the form of an autoencoder or variational autoencoders, to learn the best transformation, quantization, entropy models, and prediction methods from data. Recently, researchers have proposed an end-to-end video compression method \cite{agustsson2020scale}, using scale-space flow and scale-space warping as a generalization of flow and bilinear warping in the motion compensation step. These compression methods are either tailored for natural scenes other than biomedical data or training data dependent. Differently, our approach is designed for compact and high fidelity data fitting for diverse biomedical data.

\vspace{1mm}
\noindent\textbf{Implicit neural representation.}\quad INR is an emerging revolutionary representation of visual information, including shape \cite{genova2019deep,genova2019learning}, image \cite{mordvintsev2018differentiable,radford2015unsupervised,thies2019deferred}, video \cite{chen2021nerv,xian2021space}, and natural scenes \cite{chabra2020deep,acorn,nerf,sitzmann2019scene}, etc. The representation capability of INR can be improved from proper design of network structure \cite{chen2021nerv,mfn,lu2021compressive}, selection of activation function \cite{klocek2019hypernetwork,mehta2021modulated,siren}, embedding \cite{benbarka2021seeing,nerf,ffn,zhong2019reconstructing}, and learning strategies such as manifold learning \cite{du2021learning,mehta2021modulated,siren,sitzmann2019scene}, meta learning \cite{bergman2021fast,du2021learning,tancik2021learned} and ensemble learning \cite{aftab2021multi,kadarvish2021ensemble,niemeyer2021giraffe}. Similar to the conventional grid based discrete representations, one has to trade off between the number of neural network parameters and its representation capability. Therefore, visual data compression under the INR representation is inherently to pursue the optimal representation capability under given limited number of parameter. The proposed \M~in this paper bridges the gap between the INR and data compression, and can leverage the latest advances in the INR research to raise the compression ratio significantly. 

\vspace{1mm}
\noindent\textbf{Implicit neural compression.}\quad Recently INR has also been used for data compression and achieved promising results \cite{chen2021nerv,lu2021compressive}. They fit the original discrete grid-based data with continuous implicit functions in a transform-coding-like manner, which takes advantage of the powerful representation capabilities of neural networks while avoiding the generalization problem of data-driven. 
These primary studies demonstrate the practicability of data compression under the INR description, but the potential has far from been fully explored. The work in this paper proposes a coding framework capable of fitting diverse visual data adaptively, and is a big step towards extensive exploitation of INR's advantageous in visual data compression.

\section{Spectrum concentration property of INR}
A deep understanding of mechanisms behind the INR's fidelity degeneration on data with broad spectrum, to the best of our knowledge, is still absent. To fill this gap, we 
offer a novel perspective which theoretically analyzes how broad spectrum degrades INR's accuracy, based on a three-layer INR architecture, which can be extended to more layers.

The first layer in INR conducts frequency mapping for the input vector $\boldsymbol{v}$, and the output $\boldsymbol{z}^{(0)}$ can be described as
\begin{align}
    \boldsymbol{z}^{(0)}=\sin\left(\boldsymbol{W}^{(0)}\boldsymbol{v}\right)=\sin(\mathbf{\Omega} \boldsymbol{v}),
\end{align}
in which $\boldsymbol{W}^{(0)}$ is the weighting matrix, and we replace $\boldsymbol{W}^{(0)}\boldsymbol{v}$ with $\mathbf{\Omega}\boldsymbol{v}$ to represent the input frequencies. 
The second layer takes the linear combination of above sinusoids at varying mapping frequencies as input, and the output of the $h$th node can be written as
\begin{align}
\boldsymbol{z}_h^{(1)}&=\sin(\boldsymbol{W}_{h}^{(1)}\sin(\mathbf{\Omega} \boldsymbol{v})), \label{56_1}
\end{align}
with $\boldsymbol{W}^{(1)}$ denoting the coefficients of higher order terms of the input frequencies, followed by a sinusoidal activation function. After using Euler formula and conducting Fourier series expansion we arrive at Eqs.~\eqref{57_1} and \eqref{59_1} respectively
\begin{align}
\boldsymbol{z}_h^{(1)}&=\text{Im}\left\{\prod\limits_{k=0}^{K-1}exp\left(j\boldsymbol{W}_{h,k}^{(1)}\sin(\mathbf{\Omega}_{k} \boldsymbol{v})\right)\right\}\label{57_1}\\
&=\text{Im}\left\{\sum_{t_{0}=-\infty}^{\infty}...\sum_{t_{K-1}=-\infty}^{\infty}\prod\limits_{k=0}^{K-1} J_{t_k}(\boldsymbol{W}_{h,k}^{(1)})\text{exp}(jt_{k}\mathbf{\Omega}_{k}\boldsymbol{v})\right\}\label{59_1},
\end{align}
in which we introduce Bessel function to exhibit the coefficients of input frequencies and their high order harmonics, and 
$t_{k}\mathbf{\Omega}_{k}\boldsymbol{v}$ represents a linear combination of integer harmonics of the input frequencies, indicating expression capability of the INR architecture for a wide range of frequencies. 
By further derivation (detailed in the supplementary note 1), one can get 
\begin{align}
\boldsymbol{z}_h^{(1)}=\sum_{t_{1},...,t_{K}=-\infty}^{\infty}\left(\prod\limits_{k=0}^{K-1} J_{t_k}\boldsymbol(\boldsymbol{W}_{h,k}^{(1)})\right)\sin\left(\sum_{k=0}^{K-1}t_{k}\mathbf{\Omega}_{k}\boldsymbol{v}\right).
\end{align}
Because the MLP's weights $\boldsymbol{W}_{h,k}^{(1)}$ statistically concentrate around 0, we can derive from the asymptotic expansion of Bessel function that the coefficients of the basic frequency is far greater than those of the high frequencies. In other words, the high order harmonics in $\boldsymbol{z}_h^{(1)}$ tend to have tinier coefficients than the input frequencies $\mathbf{\Omega}$. 
Further, the third (last) layer is again a linear combination of the $\boldsymbol{z}_h^{(1)}$
\begin{align}
    f(\boldsymbol{v})=\boldsymbol{{{w}^{(2)}}^{\top}}\boldsymbol{z}^{(1)}=\boldsymbol{{w}^{(2)}}^{\top}\sin(\boldsymbol{W}^{(1)}\sin(\mathbf{\Omega} \boldsymbol{v})). \label{output of INR}
\end{align}
On the whole, the above three-layer and even more layers INR architecture will act as an inductive bias, which focuses most energy of the output signal in a narrow band around the input frequencies $\mathbf{\Omega}$. 
We refer to this phenomenon as the spectrum concentration property of INR.

Considering such spectrum concentration property, INR is more suitable for representing data with narrow spectrum consisting of a very small number of input frequencies. In other words, complex data with very spreading spectrum would suffer from severe compression artifacts because it falls out of INR's spectrum envelop.

It's worth noting that we also tested wavelet based transforms to characterize the spectrum concentration but achieved inferior
performance. Besides, since derived in the Fourier domain, 
after partitioning we do not need
wavelet-like transforms with spatially localization properties. 

\section{\M: Spectrum concentrated implicit neural compression}
\begin{figure}[t]
\centering
\includegraphics[width=0.5\textwidth]{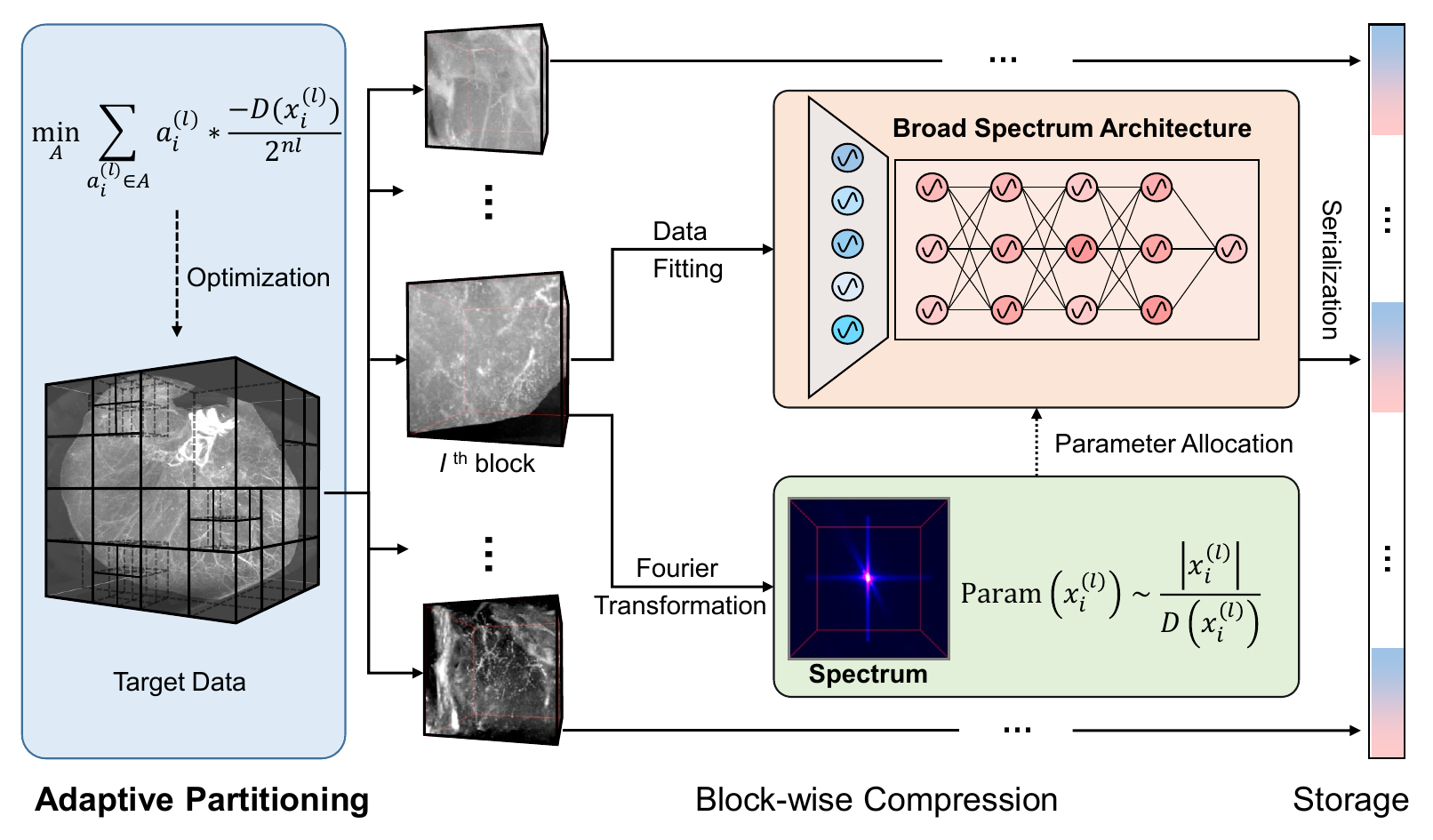}
\caption{The flow chart of the proposed \M. Firstly, the target data is adaptively partitioned into blocks (left) via solving an optimization problem, which favours a solution with all the blocks having similar spectrum coverage approximately matching the concentrated spectrum envelop of the adopted INR. Then \M~compresses each block separately with a funnel shaped neural network  architecture (middle top). The parameters are allocated according to the spectral characteristics (middle bottom) and optimized to fit the target data block (data fitting). 
Afterwards, the network parameters (mainly weights and bias) of each INR are serialized into the final compressed data (right).}
\vspace{-3mm}
\label{fig:schematic}
\end{figure}

Inspired by the above analysis, we propose \M~to boost the compression capability of INR from three perspectives, with the pipeline illustrated in Figure~\ref{fig:schematic}.
Firstly, we develop an adaptive partitioning strategy to divide the target data with far spreading spectrum distribution into spectrum concentrated blocks falling within INR's spectrum envelop. 
Secondly, we elaborately design a funnel shaped INR architecture for efficient representation of the data block. Thirdly we establish a parameter allocation strategy based on the spectral characteristics of each block.

\subsection{Adaptive partitioning}

To flexibly match the spectrum characteristics of diverse biomedical data with different complexities, we design an adaptive partitioning strategy to divide the target data 
into spectrum concentrated blocks
that can be described by only a small set of parameters. 
There are two key problems need to be addressed here: (1) How many blocks are required? (2) How to determine an optimal boundary between different blocks? Furthermore, we expected the partitioning approach to be of more advantageous features: firstly, to achieve a higher computational efficiency, the proposed scheme should be able to tackle the above two entangled problems simultaneously rather than an iterative search; secondly, the scheme can be applied for different data adaptively; besides, results can further shed light on the successive parameter allocation within each block.

Towards this goal, we implement our partitioning strategy by solving an optimization problem, which maximizes the within-block spectrum concentration. Specifically, without loss of generality, we introduced a tree data structure to provide a unified explicit representation for all possible partitioning patterns of a given data.
In an $L$ level tree data structure, 
the root node $\boldsymbol{x}\in\mathbb{R}^{S_{1}\times\cdots\times S_{N}\times c}$  represents the input data, with $S_{1}\times\cdots\times S_{N}$  denoting the spatial coordinates of $N$-dimensional data and $c$ denoting channels (e.g., color, frame, etc.). 
At finer levels, $\boldsymbol{x}^{(l)}=\{\boldsymbol{x}_{i}^{(l)}\in\mathbb{R}^{S_{1}^{l}\times\cdots\times S_{N}^{l}\times c},i=1,2,...,2^{N(l-1)}\}$ represents a set of blocks with $\boldsymbol{x}_{i}^{(l)}$  being the $i$th  node at the $l$th level, and the dimension of the block is written as
\begin{align}
S_{n}^{l}=S_{n}/{2^{N(l-1)}}, n=1,2,...,N.
\end{align}

Base on the tree-structure, a partitioning scheme can be expressed as a tree, with binary node values $a_i^{(l)} = 1$ indicating the $i$th node at level $l$ is a leaf node with sufficiently concentrated spectrum or $a_i^{(l)} = 0$ representing a non-leaf node that needs further partitioning for spectrum concentration. Then, we define an objective as Eq.~ \eqref{eq:objective} to explicitly model the within-block spectrum concentration for pursing a good partitioning scheme
\begin{align}
    \min\limits_{A}\quad\sum\limits_{a_i^{(l)}\in A}a_i^{(l)}*\frac{-D(\boldsymbol{x}_{i}^{(l)})}{2^{nl}},
    \label{eq:objective}
\end{align}
where $D(\boldsymbol{x}_{i}^{(l)})$, calculated by the ratio of the top $M$ largest values in the spectrum of $\boldsymbol{x}_{i}^{(l)}$ to the total spectrum, represents the degree of spectrum concentration of this block. $S(\boldsymbol{x}_{i}^{(l)})$ denotes the spectrum of $\boldsymbol{x}_{i}^{(l)}$
\begin{align}
    D(\boldsymbol{x}_{i}^{(l)}) = \frac{\sum\limits_{m=1}^{M}{\max(S(\boldsymbol{x}_{i}^{(l)}),m)}}{\sum S(\boldsymbol{x}_{i}^{(l)})},
\end{align}
and $A$ denotes partitioning strategies
\begin{align}
    A:=\bigcup\limits_{l=1}^{L}\{a_i^{(l)},i=1,\dots,2^{n(l-1)}\}. \label{partitioning strategies}
\end{align}
To avoid exhaustive searching, we define the feasible set of our partitioning strategy for fast optimization
\begin{align}
    \sum\limits_{a\in A}a&\le a_{max}, \label{eq:constraint1}\\ 
    \sum\limits_{a\in P(a_j^{(L)})} a&=1,j=1,\dots,2^{n(L-1)},\label{eq:constraint2}
\end{align}
where $P(\cdot)$ represents a path to the root node. Eq.~\eqref{eq:constraint1} defines the upper bound of number of blocks, limited by a theoretical minimum number of parameters in MLP to maintain representation capability \cite{trenn2008multilayer}, and Eq.~ \eqref{eq:constraint2} is introduced following the tree data structure.

In this way, partitioning strategy is formulated into an ILP (Integer Linear Programming) problem which can be solved efficiently by state-of-the-art solvers. This partitioning is a universal one being able to adaptively determine the optimal number and boundary of blocks, based on the features of the target visual data and under the given compression ratio.

\subsection{Funnel shaped neural network architecture}
After partitioning, we 
build a highly efficient INR architecture for block-wise data compression, in which the number of neurons for each layer is elaborately designed for covering the  spectrum of the target data block and achieving high representation accuracy. For the first layer conducting frequency mapping, we allocate a wide set of neurons to increase the cardinality of $\mathbf{\Omega}$. 
For the successive layers, we arrange fewer neurons at each layer to increase the depth of network. As mentioned above, $\boldsymbol{W}^{(l)}$ controls the bandwidth of the frequency modulation including translation, multiplication and harmonic expansion. Given the limited parameters it's more efficient to increase the numbers of expansion, i.e. the depth rather than the width of network, i.e. $|\boldsymbol{W}^{(l)}|$, since the former would expand the spectral support exponentially. We denote the ratio between the number of neurons in the first layer and that in the subsequent layers by $fr$
\begin{align}
    fr={|\boldsymbol{W}^{(0)}_{:,0}|}/{|\boldsymbol{W}^{(1)}_{:,0}|},
    \label{first layer ratio}
\end{align}
which is large than 1 here to take account of both the spectrum coverage and representation efficiency.

\subsection{Parameter allocation}

Adopting the optimal partitioning strategy, the within-block spectrum concentration can be subsequently calculated. 
In this section, we go one step further and develop a parameter allocation strategy based on the spectral characteristics of each block. Our strategy leverages the within-block spectrum concentration to initialize the number of parameters allocated to each block. 

Intuitively, a block with wider spectrum range (i.e., lower spectrum concentration) is more complex and deserves more parameters to maintain high reconstruction fidelity. Without loss of generality, we use an intuitive solution by setting the number of parameters allocated to each block to be proportional to its spectrum coverage. Taking $\boldsymbol{x}_{i}^{(l)}$ as an example, the parameter number linearly increases with $|\boldsymbol{x}_{i}^{(l)}|/D(\boldsymbol{x}_{i}^{(l)})$, where $|\boldsymbol{x}_{i}^{(l)}|$ denotes the size of $\boldsymbol{x}_{i}^{(l)}$.

\section{Experiments}
\vspace{2mm}
\subsection{Implementation details}
We evaluated our approach on both massive biological and medical data. For the former we used the mouse brain-wide microscopic data at sub-micrometer resolution (\textit{Neurons} and \textit{Vessels}), which have diverse brightness and structures among different regions.
For the latter we used HiP-CT dataset \cite{walsh2021imaging}, providing cellular level imaging of different organisms at multiple anatomical levels. We used all the four human organs (\textit{Lung}, \textit{Heart}, \textit{Kidney} and \textit{Brain}) contained in the dataset. 

For convenient and fair comparison, we cropped the data into the same size and used the raw data without pre-processing but 
simply normalizing the coordinates to $[- 1,1]^{3}$ in INR based methods.

For our approach, the network includes 7 layers, $fr=2.2$, the hyper parameter of sinusoidal frequency $w_0=20$, and is optimized by Adamax \cite{kingma2014adam} with learning rate of 0.001. During partitioning, we set $M=1$ and $a_{max}=50$ by default empirically.

The weights and bias of network are serialized and stored in binary, and some necessary information for decompression are stored in a small YAML file, including the network structure, original data bit depth, image size, inverse normalization basic information, etc. Due to the fact that all of the items are discrete data, countable and limited in the set of their possible values, we can easily create a table listing all potential values and store the index, allowing the file to be very small.

Because both compression and decompression are conducted in block-wise manner, we have to record the block information. The position and size of each block is recorded in the name of its compressed file, e.g., for a $16\times128\times128$ block starting from $(0, 256, 0)$ is named as $``d\_0\_15-h\_256\_383-w\_0\_127”$.
In addition, the length of file name is much smaller than the limit of modern file systems.
Moreover, there might exist block artifacts in the decoded version, especially at high compression ratio, and we use the adaptive deblocking filter \cite{list2003adaptive} to address this issue.

\subsection{Benchmark methods and evaluation metrics}

The proposed method is comprehensively compared with state-of-the-art methods, which are classified into three groups: (i) widely used commercial compression techniques, including JPEG \cite{wallace1992jpeg}, H.264 \cite{wiegand2003overview}, and HEVC \cite{sullivan2012overview}; (ii) recently proposed data-driven deep learning based ones, including DVC \cite{lu2019dvc}, SGA+BB \cite{yang20improving}, and SSF~\cite{agustsson2020scale}; (iii) INR based ones, including SIREN \cite{siren}, NeRF \cite{nerf}, and NeRV \cite{chen2021nerv}. All methods were evaluated at the similar compression ratios (only INR based methods can precisely regulate the compression ratio). We denoted the real compression ratio by BPV (bits per voxel), and use the same performance metrics 
to evaluate their compression performance. We use Accuracy with threshold for binarization as evaluation metrics for biological data with sparse structures, while PSNR and SSIM for medical data including rich details.

We used off-the-shelf implementations to evaluate the widely used commercial compression methods. JPEG was tested with its OpenCV implementation. H.264 and HEVC were tested on FFmpeg with specified bit rate settings and the applicable highest bit depth, i.e., 10-bit for H.264 and 12-bit for HEVC. 
In terms of data-driven methods, i.e. DVC, SGA+BB, and SSF, we fine-tuned the provided pre-trained models on 300 volumes ($16\times256\times256$ voxels) randomly selected from the dataset.
For SGA+BB, we used the setting with the highest performance mentioned in the original paper.
For DVC, we used the Pytorch implementation, and the ``I Frames" were compressed using JPEG method (with quality parameter set to 50).
For SSF, we adopted the CompressAI's implementation and adjusted the compression ratio by setting its quality parameters.
For INR based compression, SIREN and NeRF were both equipped with a 7-layer network structure and $fr=1$.
For SIREN, its hyper parameter $w_0\in[10,40]$.
For NeRF, the hyper parameter ``frequency" was set to 10. For NeRV, we control the compression ratio by adjusting the number and width of the fully connected layers.
Note that for 2D image compression methods, i.e. JPEG and SGA+BB, we firstly flattened our volumetric data before compression.

\begin{figure*}[t]
	\centering
	\subfigure[]{
		\includegraphics[width=0.24\textwidth]{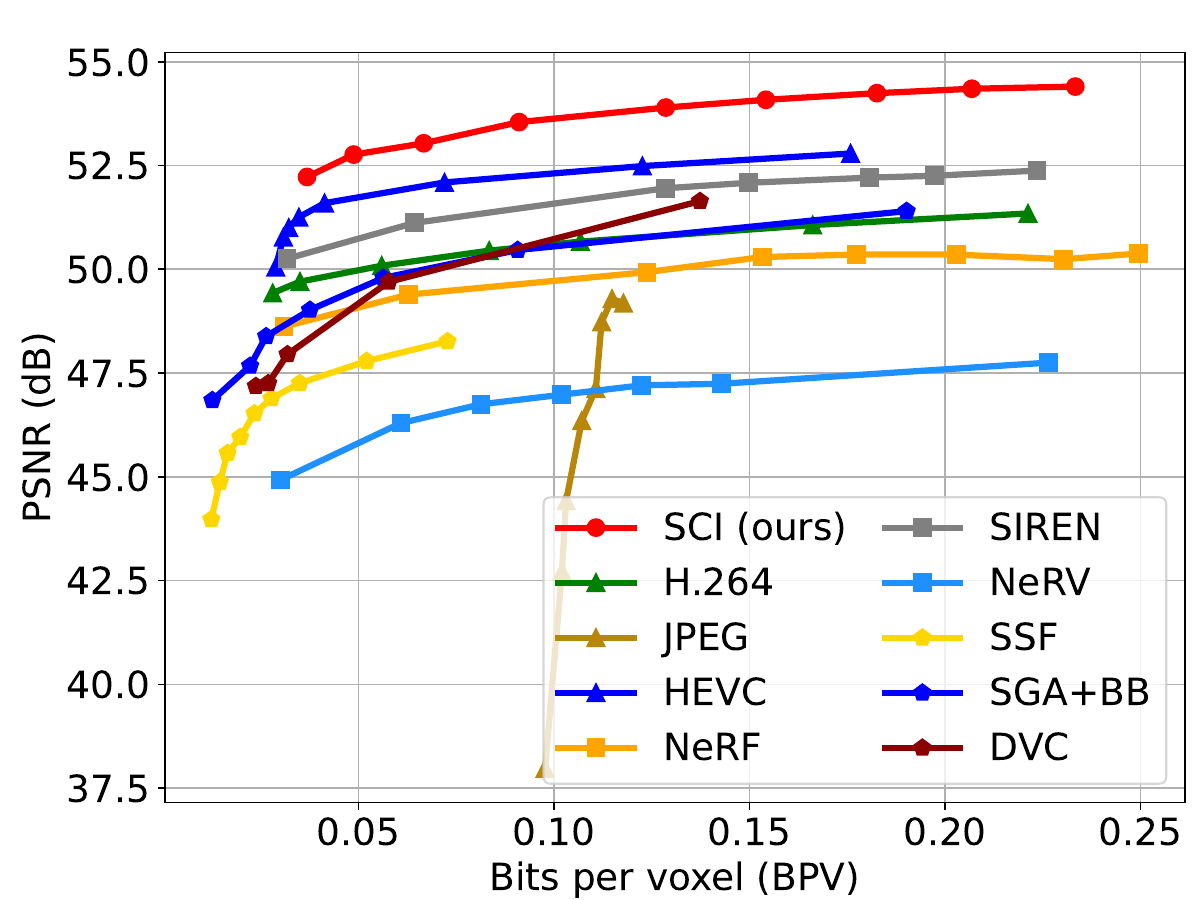}
		\includegraphics[width=0.24\textwidth]{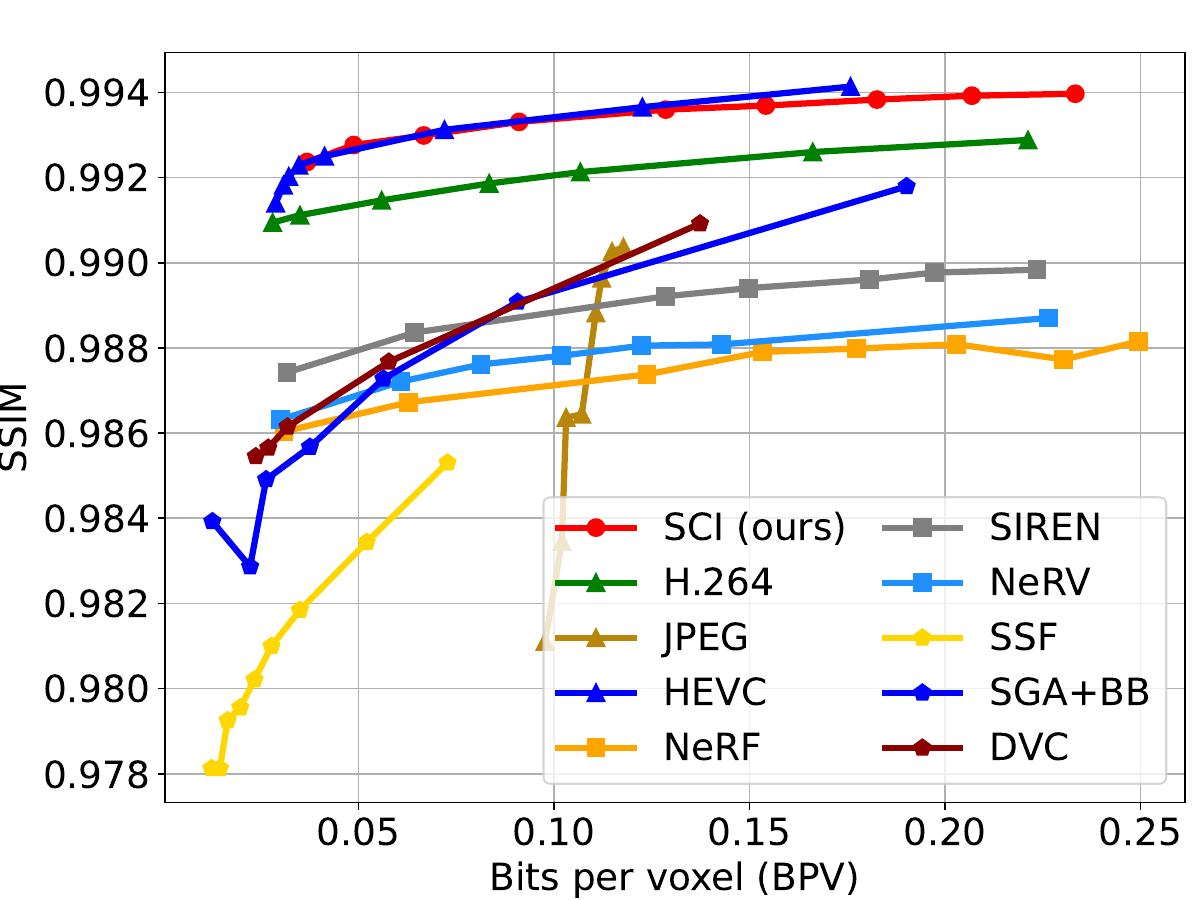}}
	\subfigure[]{
		\includegraphics[width=0.24\textwidth]{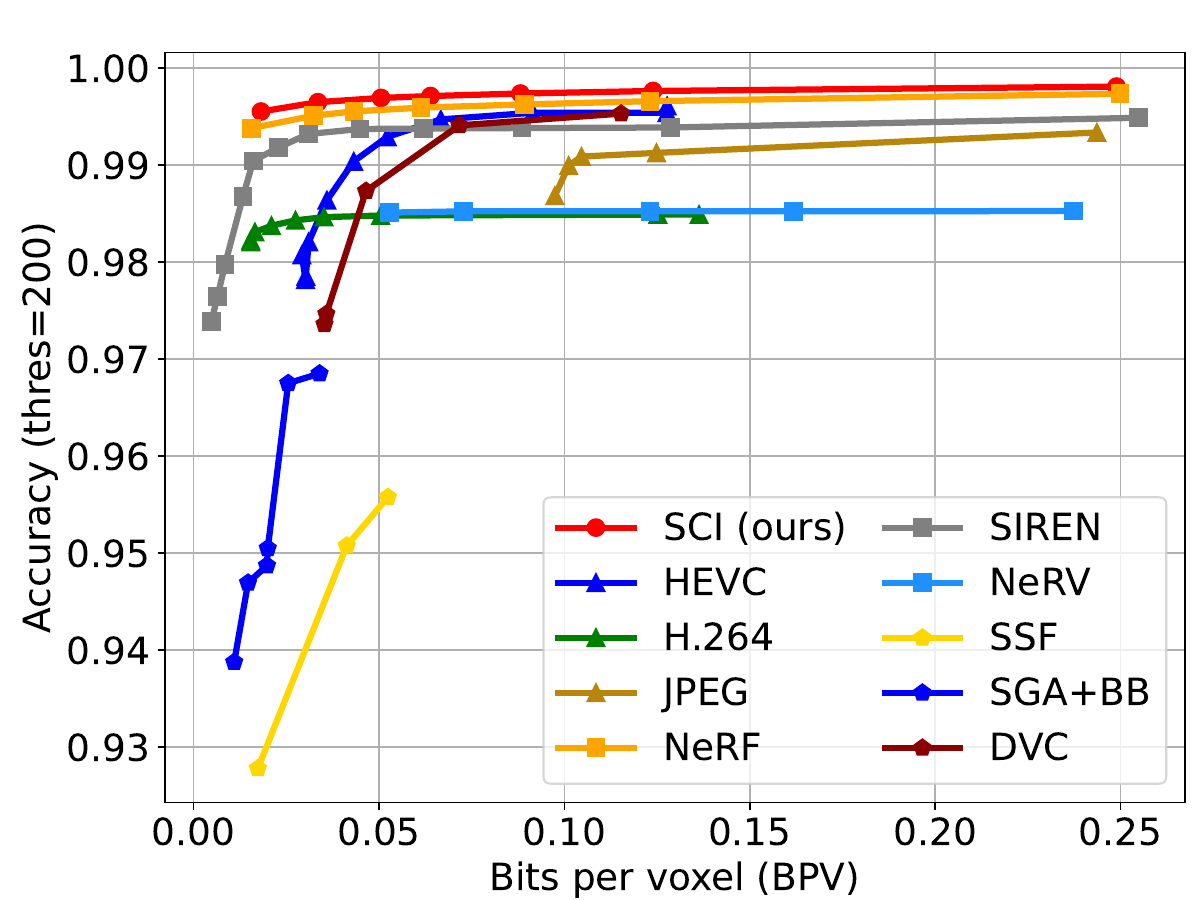}
		\includegraphics[width=0.24\textwidth]{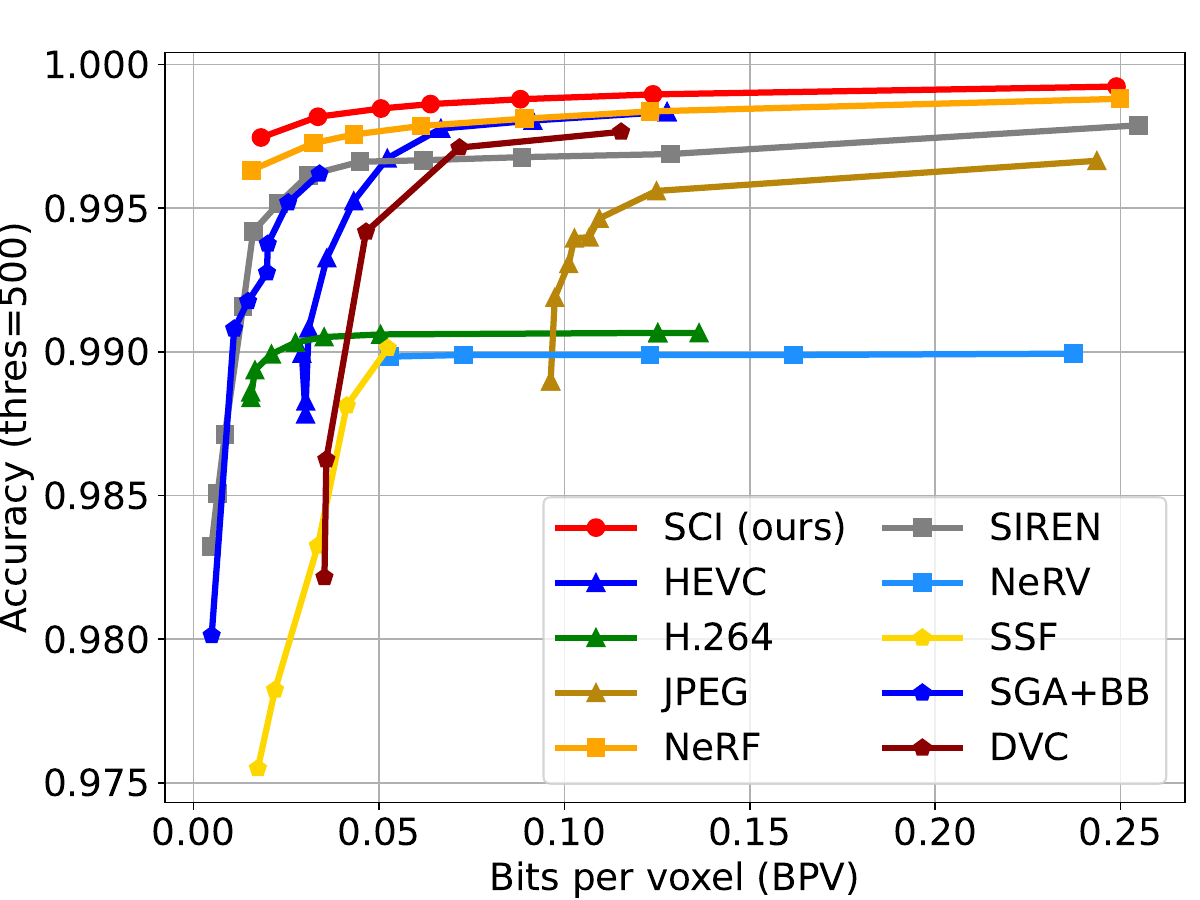}}
		\vspace{-2mm}
	\caption{Comparison of different compression methods on medical (a) and biological data (b) at different BPVs.}
	\vspace{-2mm}
	\label{fig:Comparison}
\end{figure*}

The average performance of each method on different compression ratios is plotted as a rate-distortion curve in Figure~\ref{fig:Comparison}. It is worth noting that our method outperforms state-of-the-art methods at almost all BPVs, 
showing its great capability at varying compression ratios on both biological and medical data.

\begin{table*}[t]
\centering
\small
    \caption{The average metrics on each dataset as BPV ranges from 0.01 to 0.25, with the \textcolor{red}{best} and the \textcolor{blue}{second best} marked by color. The metrics include PSNR(dB), SSIM, and Accuracy($\tau$) with $\tau$ acts as a threshold for binarization.
    }
    \setlength{\tabcolsep}{0.8mm}{
    {\tabfontsize
    \begin{tabular}{c||cc|cc|cc|cc||cc|cc} 
    \hline
        &
        \multicolumn{2}{c|}{\textit{Lung}} & 
        \multicolumn{2}{c|}{\textit{Heart}} & 
        \multicolumn{2}{c|}{\textit{Kidney}} & 
        \multicolumn{2}{c||}{\textit{Brain}} &
        \multicolumn{2}{c|}{\textit{Neurons}} &
        \multicolumn{2}{c}{\textit{Vessels}} \\
        \cline{2-13}
        Method 
        & {\scriptsize ~~~PSNR~~~} & {\scriptsize ~~~SSIM~~~}
        & {\scriptsize ~~~PSNR~~~} & {\scriptsize ~~~SSIM~~~}
        & {\scriptsize ~~~PSNR~~~} & {\scriptsize ~~~SSIM~~~}
        & {\scriptsize ~~~PSNR~~~} & {\scriptsize ~~~SSIM~~~} 
        & {\scriptsize ~~~Acc. (200)~~~} & {\scriptsize ~~~Acc. (500)~~~} 
        & {\scriptsize ~~~Acc. (200)~~~} & {\scriptsize ~~~Acc. (500)~~~} 
        \\ 
    \hline
        \M~(ours) & \textcolor{blue}{45.63} & \textcolor{blue}{0.9734} & \textcolor{red}{63.23} & \textcolor{red}{0.9995} & \textcolor{red}{50.15} & \textcolor{blue}{0.9896} & \textcolor{red}{50.82} & \textcolor{blue}{0.9927} &
        \textcolor{red}{0.9986} &  
        \textcolor{red}{0.9995} &
        \textcolor{red}{0.9944} &   
        \textcolor{red}{0.9970} \\ 
    \hline
        JPEG & 42.32 & 0.9599 & 48.20 & 0.9970 & 44.76 & 0.9850 & 43.90 & 0.9757 & 
        0.9949 & 0.9969 &
        0.9846 & 0.9866
        \\ 
    \hline
        H.264 & 45.30 & 0.9733 & 57.66 & 0.9990 & 49.59 & 0.9894 & 50.08 & 0.9925 & 
        0.9950 & 0.9975 &
        0.9710 & 0.9797 
        \\  
    \hline
        HEVC & \textcolor{red}{46.15} & \textcolor{red}{0.9763} & \textcolor{blue}{61.54} & \textcolor{blue}{0.9992} & \textcolor{blue}{50.09} & \textcolor{red}{0.9902} & \textcolor{blue}{50.31} & \textcolor{red}{0.9932} & 
        0.9947 & 0.9976 &
        0.9848 & 0.9891  
        \\ 
    \hline
        SIREN & 45.48 & 0.9727 & 55.49 & 0.9989 & 48.99 & 0.9891 & 49.22 & 0.9905 & 
        \textcolor{blue}{0.9983} & 
        \textcolor{blue}{0.9993} &
        0.9866 & 0.9918
        \\ 
    \hline
        NeRF & 44.56 & 0.9687 & 57.77 & 0.9991 & 49.77 & 0.9890 & 49.77 & 0.9919 & 
        0.9980 & 0.9992 &
        \textcolor{blue}{0.9913} & 
        \textcolor{blue}{0.9949}  
        \\ 
    \hline
        NeRV & 43.36 & 0.9681 & 53.67 & 0.9982 & 46.71 & 0.9887 & 46.83 & 0.9908 & 
        0.9948 & 0.9970 &
        0.9756 & 0.9828
        \\ 
    \hline
        DVC & 44.42 & 0.9698 & 54.36 & 0.9987 & 48.75 & 0.9887 & 47.19 & 0.9887 & 
        0.9935 & 0.9970 &
        0.9828 & 0.9901
        \\ 
    \hline
        SGA+BB & 44.95 & 0.9700 & 54.38 & 0.9977 & 48.51 & 0.9875 & 48.68 & 0.9887 & 
        0.9920 & 0.9957 &
        0.9620 & 0.9803 
        \\ 
    \hline
        SSF & 43.12 & 0.9643 & 52.98 & 0.9980 & 46.86 & 0.9865 & 46.92 & 0.9901 & 
        0.9850 & 0.9951 &
        0.9504 & 0.9525
        \\ 
    \hline
    \end{tabular}}
    }
    \label{Table:1}
\end{table*}

\begin{figure}[t]
\centering
\includegraphics[width=\linewidth]{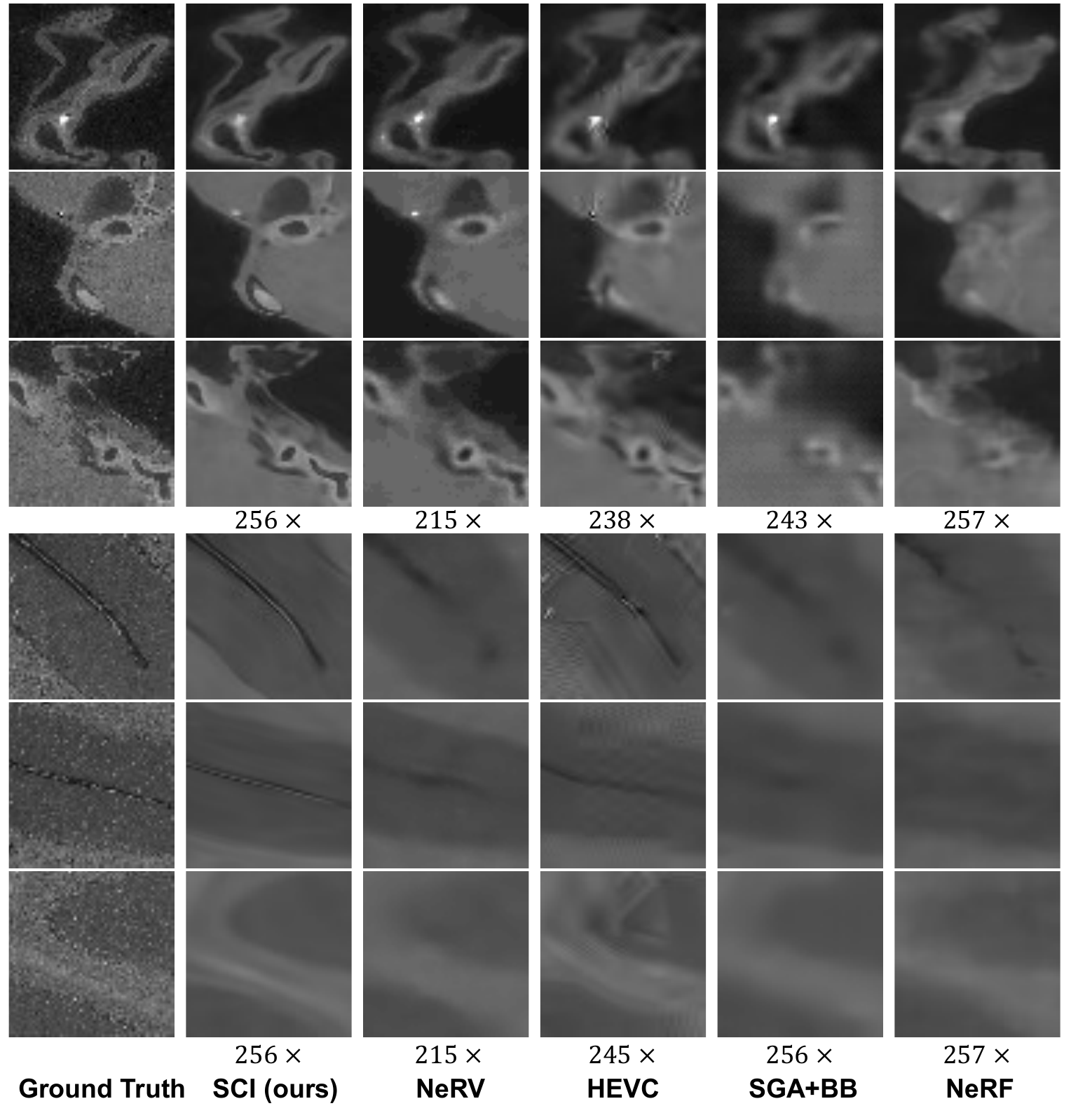}
\caption{Visual comparisons with state-of-the-arts on medical data---2D slices from 3D volumes. Rows 1-3 show the regions-of-interest (ROIs) of tubules on kidney. Rows 4-6 show the ROIs of white and grey matters on brain.
The compression ratios are labeled at bottom.}
\vspace{-6mm}
\label{fig:compare_details}
\end{figure}

For more detailed analysis, we summarize the scores on each dataset in Table~\ref{Table:1}. 
Since only the INR based method can control the BPV precisely, the others fluctuate around the specified BPV, we calculate the average performance of all the algorithms within the intersection of their BPVs range on each dataset.
Generally, our method and HEVC rank top two on all organs from CT. Since our method is a lossy compression and tends to suppress noise, on noisier data such as \textit{Lung} and \textit{Kidney}, our method is slightly inferior to HEVC on SSIM, while on cleaner data such as \textit{Heart} and \textit{Brain}, the PSNR of our method surpasses HEVC by 1.69dB and 0.51dB respectively. For \textit{Neurons} and \textit{Vessels}, the INR based methods have significant advantages and ours works best.

It should be noted that unlike the compression of natural images, the PSNR is generally high for medical data. Therefore, even though the PSNR is higher than 40dB, it does not mean that there is no distortion in the visual effect, which is closely related to the type of target data.

Some visual comparisons of decompressed organ slices are reported in Figure~\ref{fig:compare_details} (see Supplementary Figures~\ref{fig_sup: compare_details_kidney_sup},\ref{fig_sup: compare_details_brain_sup}, and~\ref{fig_sup: compare_details_neurons_sup} for more results). We did the same post-processing on the decompressed data for better visual comparison.
At similar compression ratio around $256\times$(since SGA+BB and HEVC cannot precisely regulate compression ratio), we consistently outperform existing INR based (NeRV and NeRF), data-driven deep learning (SGA+BB) and commercial compression methods (HEVC).
NeRF produces blurry artifacts on most of informative regions, since the broad spectrum of the target data goes beyond the spectrum envelop of the network. 
NeRV retains the general structures but suffers from the block artifacts caused by the aliasing in convolution. 
In the depressed results of SGA+BB, one can notice checkerboard artifacts in the regions with rich details.
HEVC is able to capture details in ``I Frames" well but there exist severe block artifacts in ``P Frames".
This can distort the edges in the volumes, such as the tubules in Figure~\ref{fig:compare_details}'s rows 1-3 and white and grey matters in rows 4-6, 
which would be unusable for medical research as well as other downstream tasks. 
In addition, HEVC struggles with the ghosting artifacts from neighbor slices in ``P Frames", causing position offset in some regions,
which would result in discontinuous 3D structure, after decompression, leading to wrong medical diagnosis or incorrect results in downstream tasks. On the contrary, the proposed approach is of the highest fidelity after decompression. We also compared the speed of these algorithms as shown in Supplementary Table~\ref{tab: speed}.  Our method is slower in compression stage but can decompress faster.

\subsection{Ablation studies}
We validated our adaptive partitioning, network architecture design, and parameter allocation strategy via an extensive ablation study, with the results shown in Table~\ref{Table:2} (see Supplementary Table.\ref{Table_sup: 1} for detailed results on each dataset). The first row shows the performance of our complete model as a reference. The complete model adopted adaptive partitioning, funnel shaped network architecture (i.e., more neurons in the first layer than in successive layers), and spectrum-width proportional parameter allocation. 
In rows 2-8, we removed one of these three components at a time from the complete model to quantitatively evaluate their contribution to the final performance.

\begin{table}[h]
\centering
\small
    \caption{Results of ablation studies on medical and biological data.}
    {\tabfontsize
    \begin{tabular}{l||cc||cc} \hline &
        \multicolumn{2}{c||}{~\textit{Medical data}} & 
        \multicolumn{2}{c}{~\textit{Biological data}} \\
        \cline{2-5}
        Method 
        & {\scriptsize ~~~PSNR~~~} & ~{\scriptsize ~~~SSIM~~~}
        & {\scriptsize ~~~Acc. (200)~~~} & ~{\scriptsize ~~~Acc. (500)~~~}
        \\
    \hline
        \M~(ours) & \textbf{49.29} & \textbf{0.9558} & \textbf{0.9752} & \textbf{0.9865} \\ 
    \hline
        EP & 48.67 & 0.9546 & 0.9706 & 0.9842 \\ 
        NP & 47.76 & 0.9532 & 0.9670 & 0.9835 \\ 
    \hline
        $fr=1.0$ & 49.24 & 0.9556 & 0.9537 & 0.9738 \\ 
        $fr<1.0$ & 48.87 & 0.9540 & 0.9354 & 0.9600 \\ 
    \hline
        By Size & 49.01 & 0.9560 & 0.9708 & 0.9849 \\ 
        By $D$ & 48.82 & 0.9558 & 0.9745 & 0.9861 \\ 
        Equal & 48.35 & 0.9554 & 0.9707 & 0.9855 \\ 
    \hline
    \end{tabular}}
    \label{Table:2}
\end{table}

In rows 2-3, we removed adaptive partitioning and adopted either naive equidistant partitioning (EP) or no partitioning (NP). The comparison between reconstruction fidelity of models with adaptive partitioning and without partitioning under various BPVs on medical data are shown in Figure~\ref{Fig:3} (see Supplementary Figure.~\ref{fig_sup: Comparison between the reconstruction fidelity of our compressor with adaptive partitioning and without partitioning on each dataset} for detailed results on all the datasets). Applying adaptive partitioning according to the spectrum concentration of data tend to achieve higher reconstruction fidelity than that without partitioning, which indicates the necessity of adaptive partitioning. It is worth noting that the performance of no-partitioning strategy decreases as BPV increases, since the neural network is too large to optimize. In all the cases, adaptive partitioning is consistently better than equidistant partitioning. 

Rows 4–5 show how the performance decreases as $fr$ reduces. One can see that for almost all the test data, the reconstruction fidelity  consistently gets higher when using a funnel shaped architecture with larger opening, i.e., wider spectrum coverage. Experimentally, the optimal ratio $fr$ varies with the target data and compression ratio.

In our approach, the number of parameters of each block is proportional to $|\boldsymbol{x}_{i}^{(l)}|/D(\boldsymbol{x}_{i}^{(l)})$. In rows 6-8, we compared our parameter allocation scheme with three different adhoc counterparts: the number of  allocated parameters in rows 6-7 is proportional to $|\boldsymbol{x}_{i}^{(l)}|$ and $1/D(\boldsymbol{x}_{i}^{(l)})$ respectively; the 8th row shows the result by equally allocating parameters among data blocks. Compared with the last three rows we can conclude that  allocating parameters based on the concentration of data spectrum indeed can improve the compression performance.

Summarizing all the ablation studies, three strategies all contribute to the final high performance, and the improvement is consistent over all the test data. Comparatively, the adaptive partitioning brings the largest quantitative benefit.

\begin{figure}[t]
    \vspace{-2mm}
	\centering
	\includegraphics[width=0.232\textwidth]{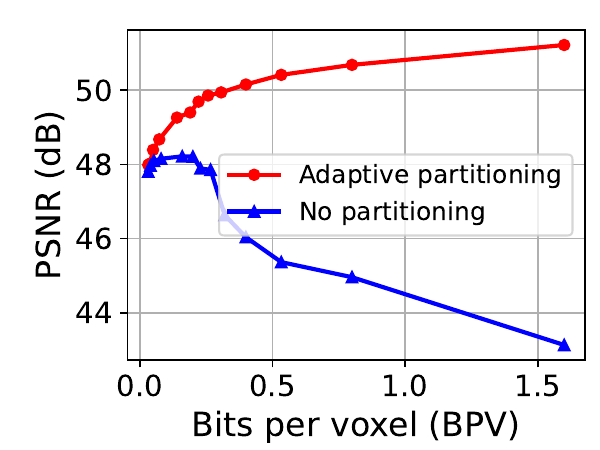}
	\includegraphics[width=0.232\textwidth]{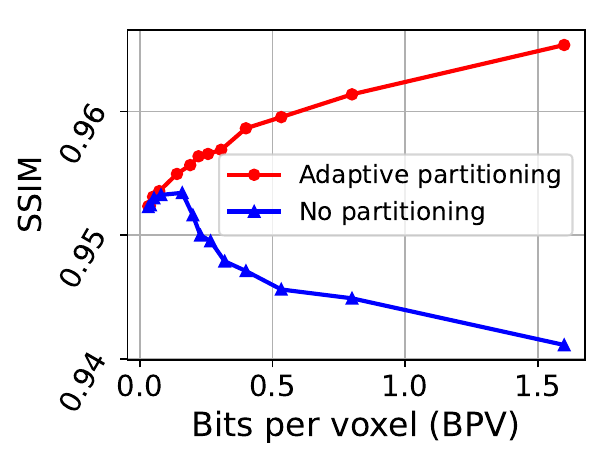}
	\caption{Comparison between SCI's reconstruction fidelity with adaptive partitioning and without partitioning on medical data, in terms of PSNR and SSIM.}
	\vspace{-6mm}
	\label{Fig:3} 
\end{figure}

In addition, the reasonableness of using $D(\boldsymbol{x}_{i}^{(l)})$ to express the data spectrum concentration is demonstrated by the experimental results in the Supplementary Figure.~\ref{fig_sup: Statistical graph of the relationship between reconstruction fidelity  and the degree of spectrum concentration}.
\section{Conclusions}

In this paper we report a general INR based compression approach for massive diverse biomedical data. The proposed approach is built on a mathematical analysis of the INR’s spectrum  concentration property, and theoretical exploration on the mechanisms behind the performance variation over diverse data with different spectrum coverages. 
Based on the analytical insight, we proposed an adaptive partitioning strategy for addressing spectrum-spreading data, an elaborately designed INR architecture for efficient data compression, 
and a parameter allocation strategy for further performance boosting. 
Experiments indicate that our method outperforms state-of-the-art methods including conventional (JPEG, H.264, HEVC), data-driven (DVC, SGA+BB, SSF), and existing INR based ones (SIREN, NeRF, NeRV) on a large diversity of biological and medical data. Our method inherits the benefit of INR, such as the ability to precisely regulate the compression ratio and the freedom to tailor fidelity of different local regions.
This paper is a pioneering study, which will pave the way for compression on the emerging neural representation.

\vspace{1mm}
\textbf{Limitations.~~~~}As an emerging technique, our approach is limited in several aspects. Firstly, current implementation requires relatively longer time than commercial compression methods, since it takes time to search for the proper network parameters fitting the target data. Secondly, 
we only remove within-block redundancy, 
taking inter-block redundancy into consideration (similar to incorporating non-local similarity) can further improve the compression. We expect to devote more efforts on acceleration and performance boosting of this promising new compression approach.

\vspace{1mm}
\textbf{Future work.~~~~}We plan to combine meta-learning to find the manifold of a specific biomedical data in INR space to boost \M's compression speed.
It is also a worth studying topic to combine network compression for shortening the coding length and non-local ensemble for removing inter-block redundancy.

\section*{Acknowledgements}
We acknowledge MoE Key Laboratory of Biomedical Photonics (Huazhong University of Science and Technology) for sharing their mouse brain-wide microscopic data.

This work is supported by Beijing Natural Science Foundation (Grant No. Z200021) and the National Natural Science Foundation of China (Grant Nos. 61931012, 62088102).
\bibliography{references}

\newpage
\onecolumn
\setcounter{table}{0}
\setcounter{figure}{0}
\setcounter{equation}{0}
\captionsetup[figure]{name={Supplementary Figure.},labelsep=period}
\captionsetup[table]{name={Supplementary Table.},labelsep=period}
\section*{Supplementary Notes}
\subsection{1. Derivation for Eq(2)-Eq(5)}
\begin{align}
\boldsymbol{z}_h^{(1)}&=\sin(\boldsymbol{W}_{h}^{(1)}\sin(\mathbf{\Omega} \boldsymbol{v}))\\
&=\sin\left(\sum\limits_{k=0}^{K-1}\boldsymbol{W}_{h,k}^{(1)}\sin(\mathbf{\Omega}_{k} \boldsymbol{v})\right)\\
&=\text{Im}\left\{\text{exp}\left( j\left(\sum\limits_{k=0}^{K-1}\boldsymbol{W}_{h,k}^{(1)}\sin(\mathbf{\Omega}_{k} \boldsymbol{v})\right)\right)\right\}\\
&=\text{Im}\left\{\prod\limits_{k=0}^{K-1}\text{exp}\left(j\boldsymbol{W}_{h,k}^{(1)}\sin(\mathbf{\Omega}_{k} \boldsymbol{v})\right)\right\}\label{57}\\
&=\text{Im}\left\{\prod\limits_{k=0}^{K-1}\sum_{t_{k}=-\infty}^\infty J_{t_k}\boldsymbol{W}_{h,k}^{(1)}\text{exp}(jt_{k}\mathbf{\Omega}_{k} \boldsymbol{v})\right\}\label{58}\\
&=\text{Im}\left\{\sum_{t_{0}=-\infty}^{\infty}...\sum_{t_{K-1}=-\infty}^{\infty}\prod\limits_{k=0}^{K-1} J_{t_k}\boldsymbol{W}_{h,k}^{(1)}\text{exp}(jt_{k}\mathbf{\Omega}_{k} \boldsymbol{v})\right\}\label{59}\\
&=\sum_{t_{1},...,t_{K}=-\infty}^{\infty}\text{Im}\left\{\left(\prod\limits_{k=0}^{K-1} J_{t_k}\boldsymbol{W}_{h,k}^{(1)}\right)\text{exp}\left(j\sum_{k=0}^{K-1}t_{k}\mathbf{\Omega}_{k} \boldsymbol{r}\right)\right\}\\
&=\sum_{t_{1},...,t_{K}=-\infty}^{\infty}\left(\prod\limits_{k=0}^{K-1} J_{t_k}\boldsymbol{W}_{h,k}^{(1)}\right)\text{Im}\left\{\text{exp}\left(j\sum_{k=0}^{K-1}t_{k}\mathbf{\Omega}_{k} \boldsymbol{r}\right)\right\}\\
&=\sum_{t_{1},...,t_{K}=-\infty}^{\infty}\left(\prod\limits_{k=0}^{K-1} J_{t_k}\boldsymbol{W}_{h,k}^{(1)}\right)\sin\left(\sum_{k=0}^{K-1}t_{k}\mathbf{\Omega}_{k} \boldsymbol{r}\right)
\end{align}
where $J_m(\cdot)$ represents the Bessel function of the first kind of order $m$ and Eq.~(5) follows from the Fourier series expansion of $ \text{exp}(jw\sin(x))$:
\begin{align}
 \text{exp}(jw\sin(x))=\sum_{m=-\infty}^{\infty}J_{m}(w)\text{exp}(jmx).  
\end{align}

\newpage
\section*{Supplementary Figures}
\begin{figure}[H]
\centering
\includegraphics[width=\linewidth]{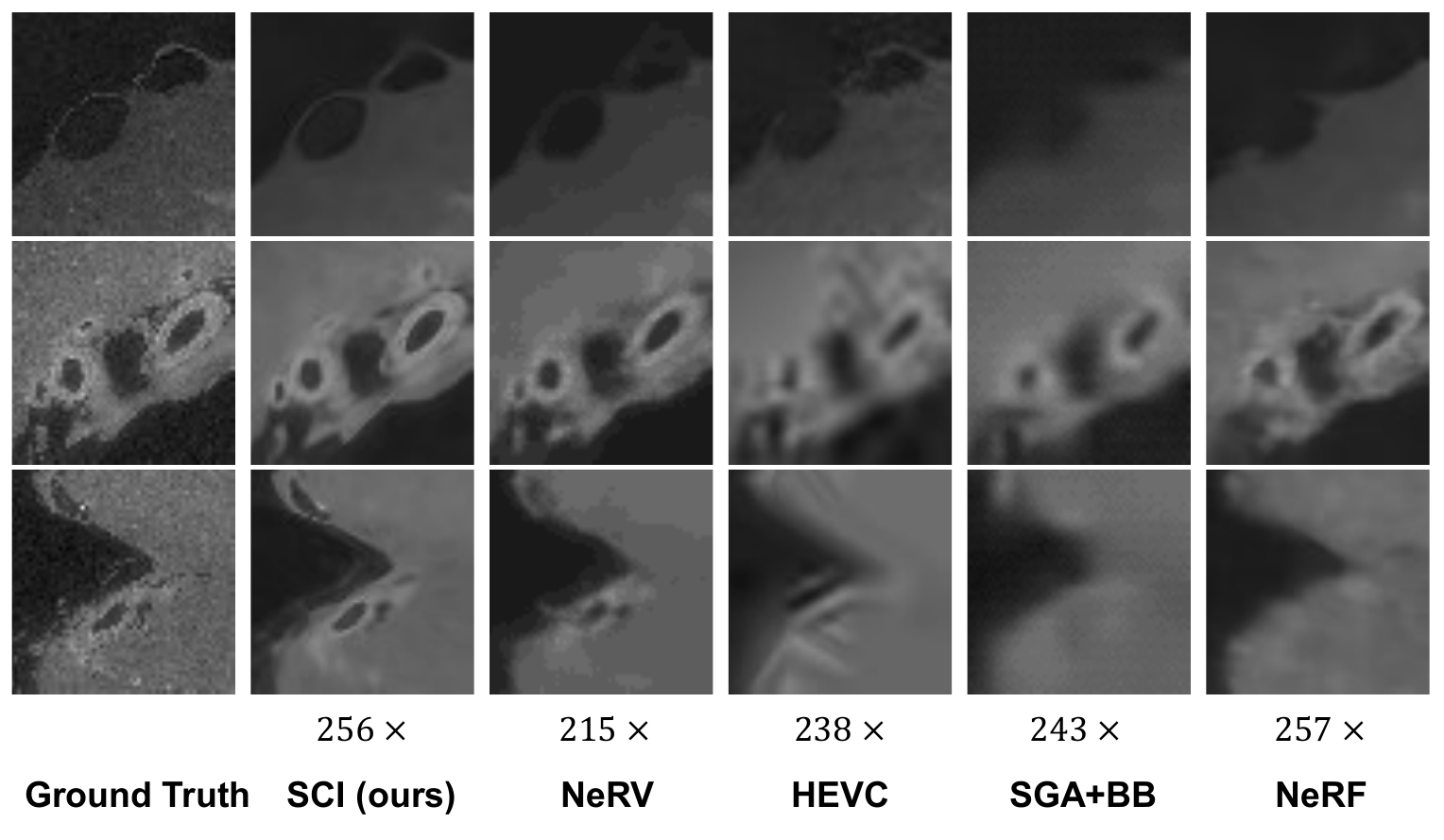}
\caption{Visual comparisons with state-of-the-arts on \textit{Kidney} dataset---2D slices from 3D volumes.
The compression ratio of each are labeled at bottom.
}
\label{fig_sup: compare_details_kidney_sup}
\end{figure}

\begin{figure}[H]
\centering
\includegraphics[width=\linewidth]{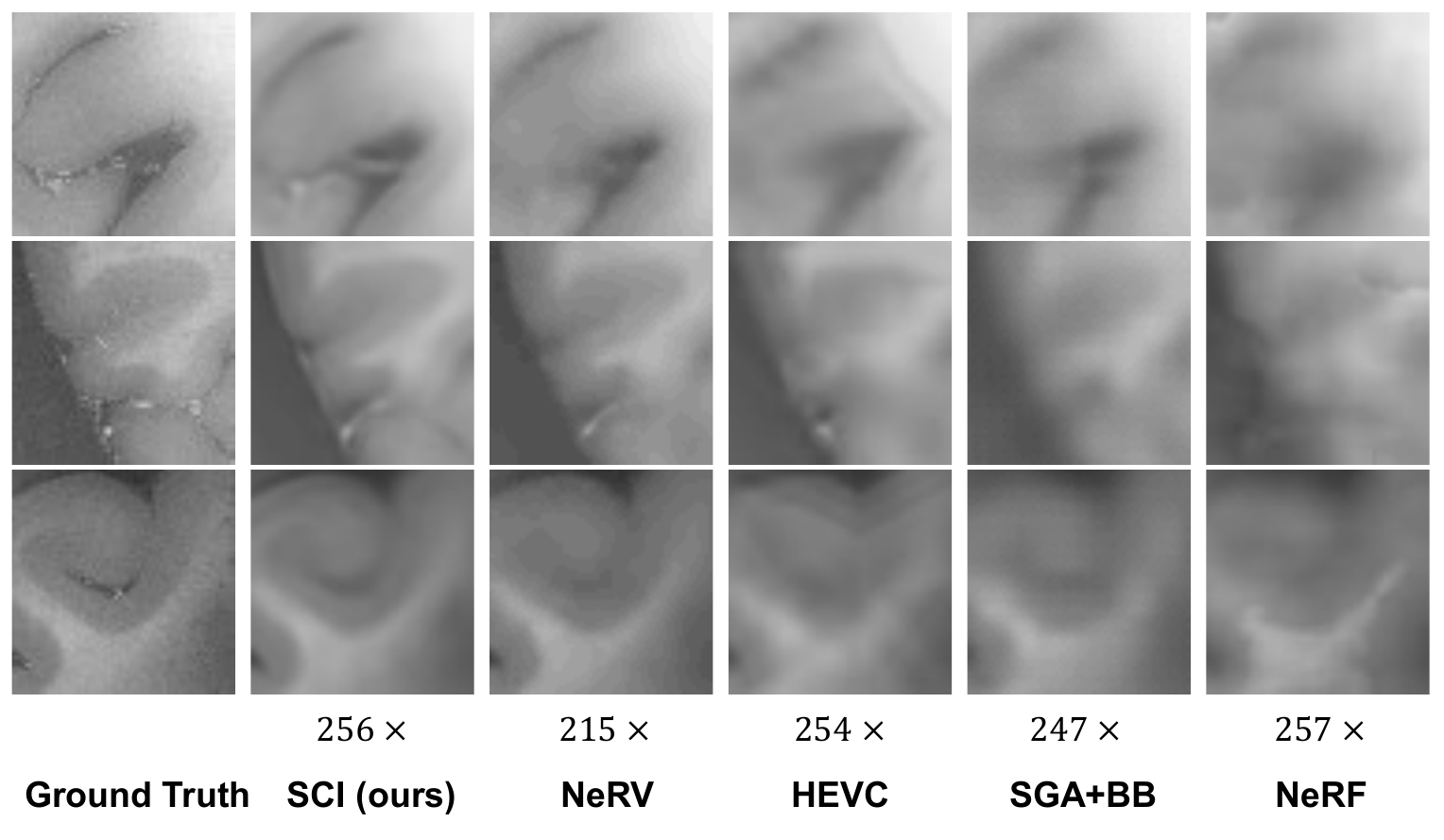}
\caption{Visual comparisons with state-of-the-arts on \textit{Brain} dataset---2D slices from 3D volumes.
The compression ratio of each are labeled at bottom.
}
\label{fig_sup: compare_details_brain_sup}
\end{figure}

\begin{figure}[H]
\centering
\includegraphics[width=\linewidth]{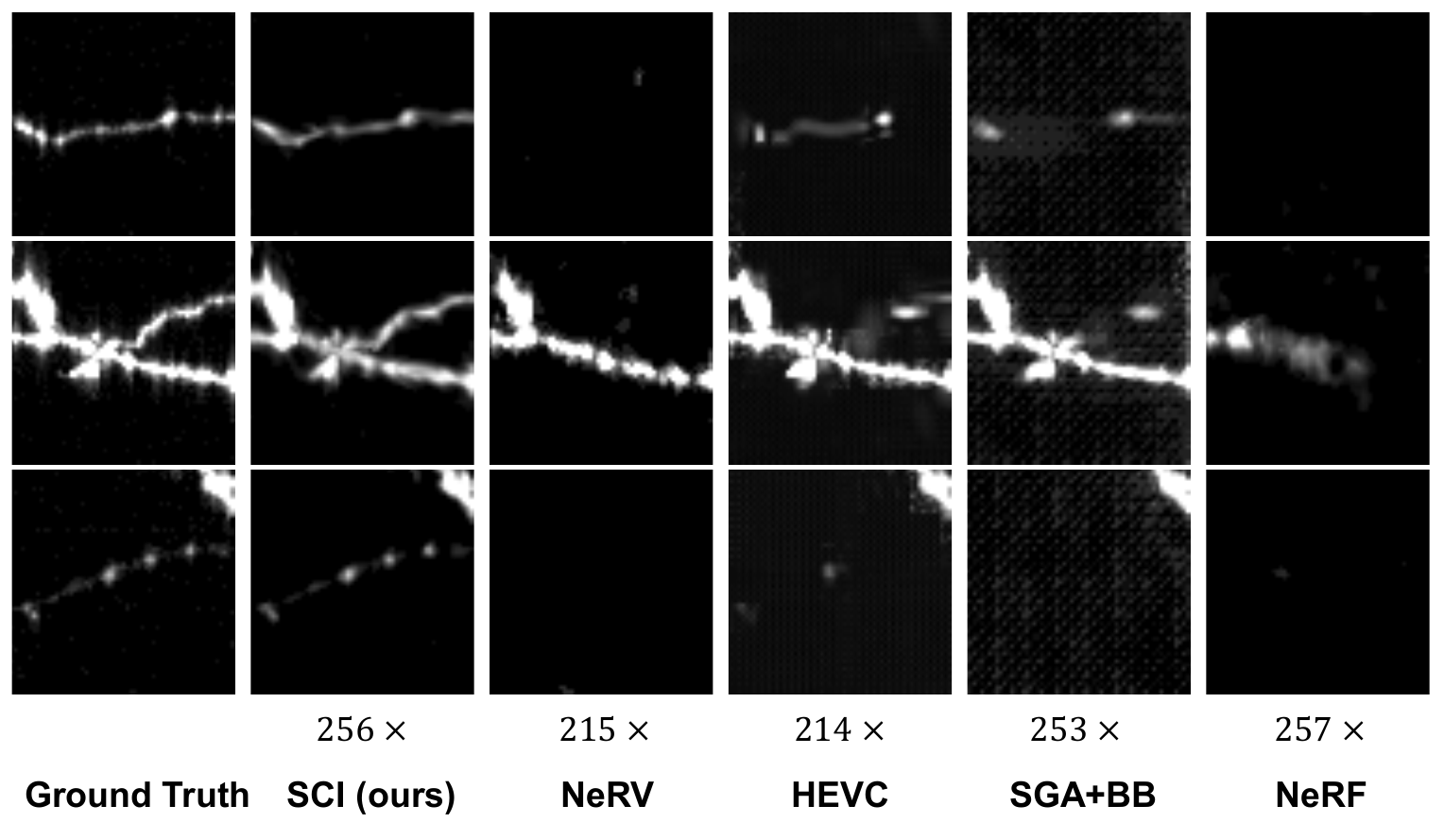}
\caption{
Visual comparisons with state-of-the-arts on \textit{Neurons} dataset---2D maximum intensity projections from 3D volumes.
The compression ratio of each are labeled at bottom.
}
\label{fig_sup: compare_details_neurons_sup}
\end{figure}

\begin{figure}[H]
	\centering
	\subfigure{
		\includegraphics[width=0.23\textwidth]{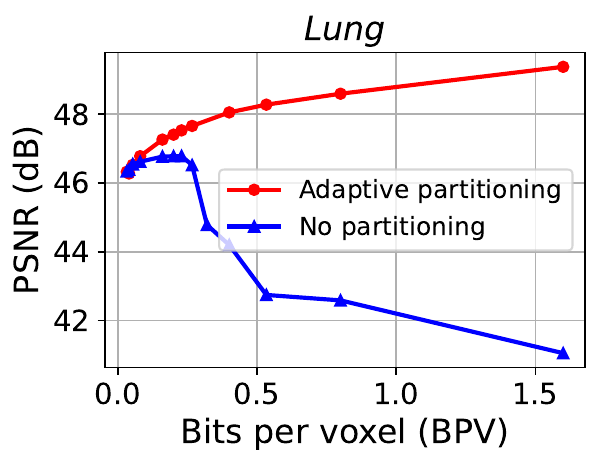}}
	\subfigure{
		\includegraphics[width=0.23\textwidth]{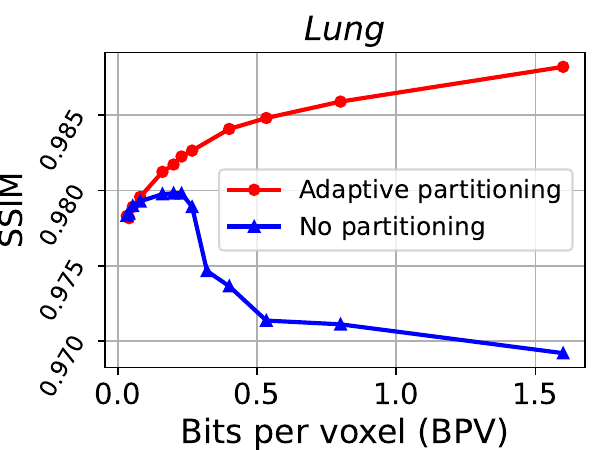}}
	\subfigure{
		\includegraphics[width=0.23\textwidth]{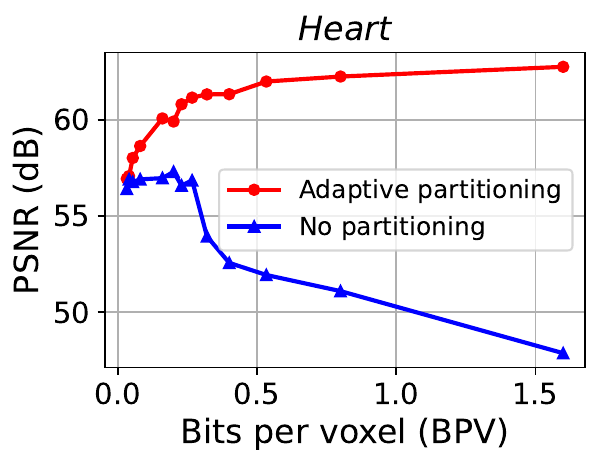}}
	\subfigure{
		\includegraphics[width=0.23\textwidth]{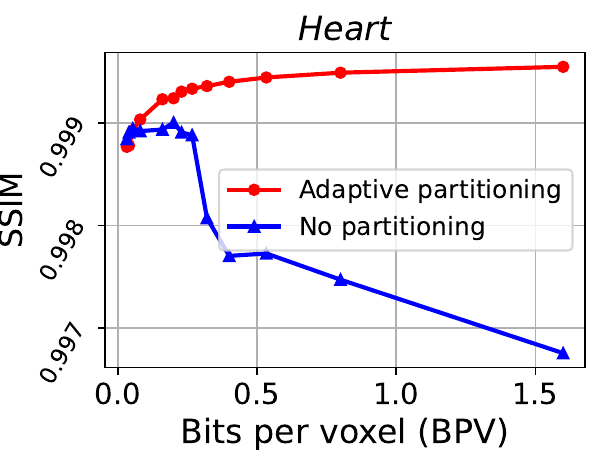}}
	\subfigure{
		\includegraphics[width=0.23\textwidth]{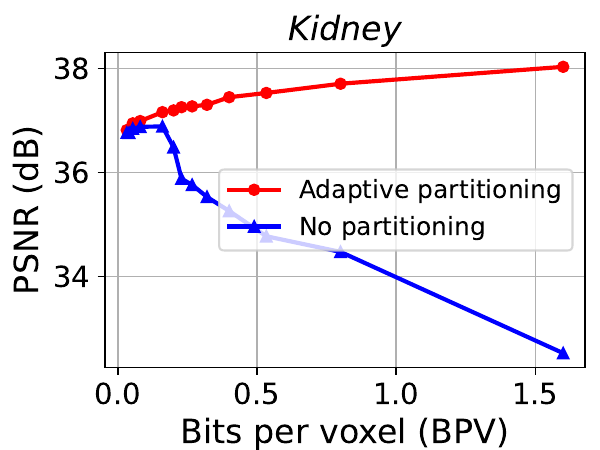}}
	\subfigure{
		\includegraphics[width=0.23\textwidth]{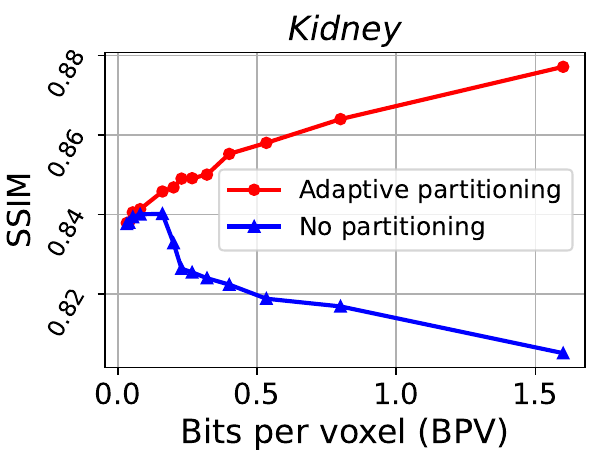}}
	\subfigure{
		\includegraphics[width=0.23\textwidth]{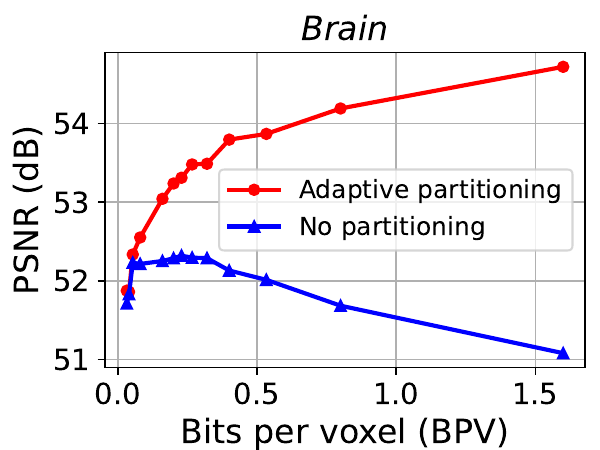}}
	\subfigure{
		\includegraphics[width=0.23\textwidth]{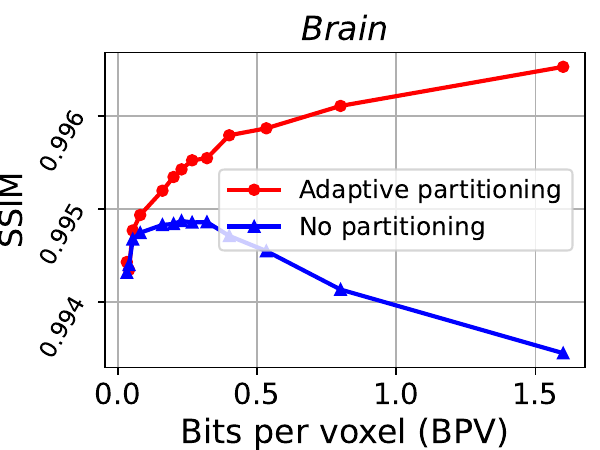}}
	\subfigure{
		\includegraphics[width=0.23\textwidth]{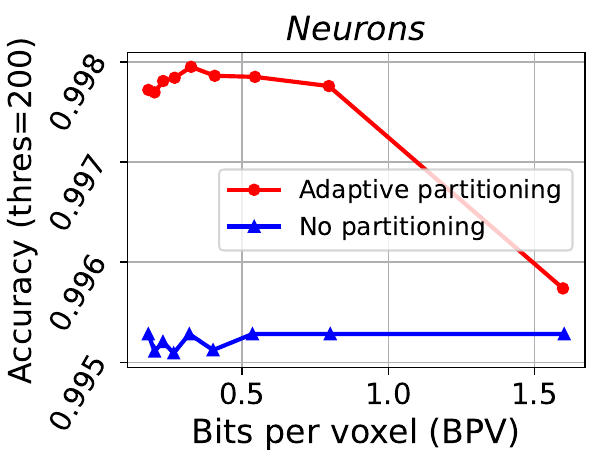}}
	\subfigure{
		\includegraphics[width=0.23\textwidth]{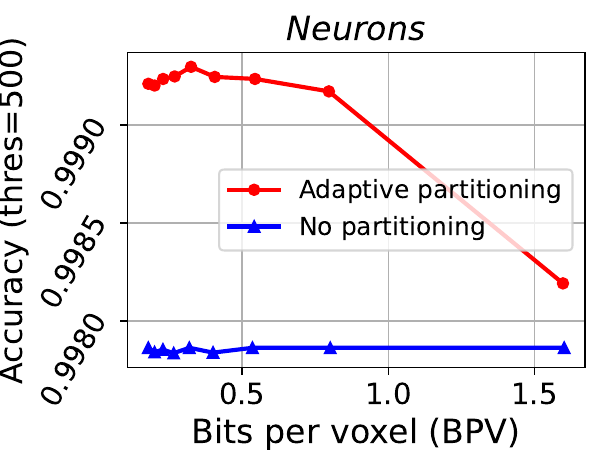}}
	\subfigure{
		\includegraphics[width=0.23\textwidth]{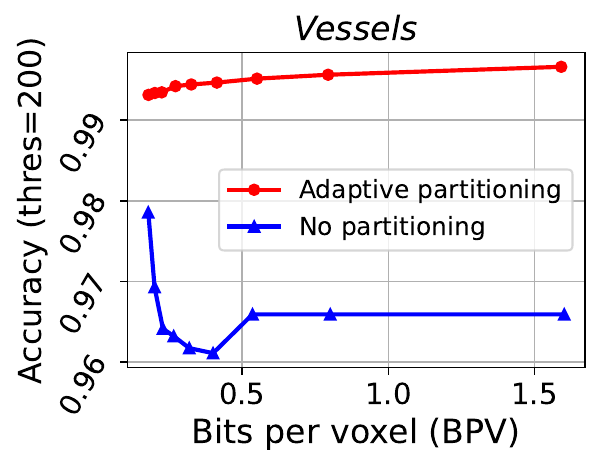}}
	\subfigure{
		\includegraphics[width=0.23\textwidth]{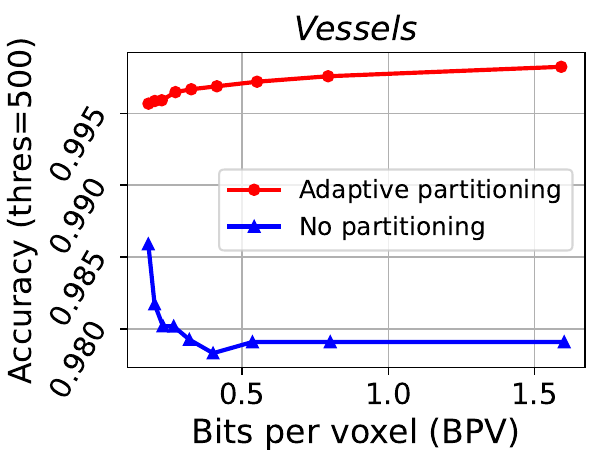}}
	\caption{Comparison between the reconstruction fidelity of our compressor with adaptive partitioning and without partitioning on each dataset.}
	\label{fig_sup: Comparison between the reconstruction fidelity of our compressor with adaptive partitioning and without partitioning on each dataset} 
\end{figure}

\begin{figure}[H]
	\centering
	\subfigure{
		\includegraphics[width=0.23\textwidth]{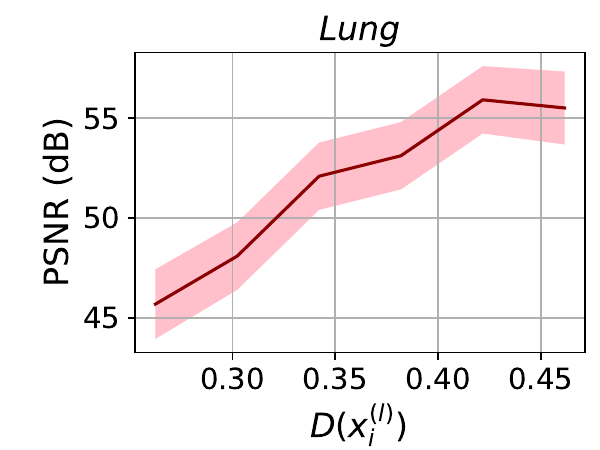}}
	\subfigure{
		\includegraphics[width=0.23\textwidth]{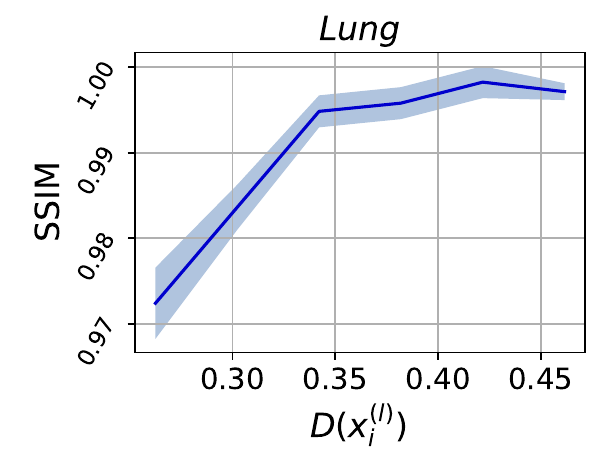}}
	\subfigure{
		\includegraphics[width=0.23\textwidth]{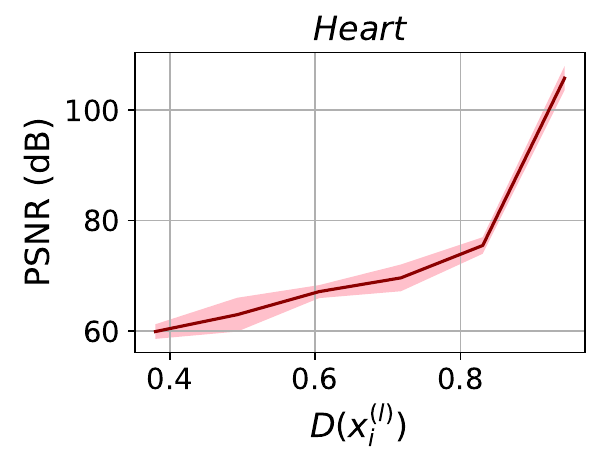}}
	\subfigure{
		\includegraphics[width=0.23\textwidth]{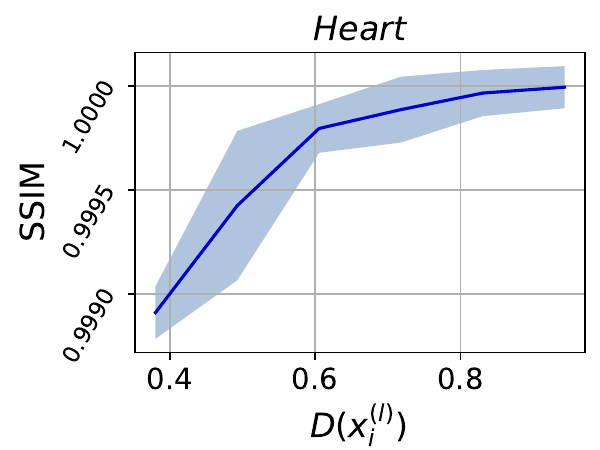}}
	\subfigure{
		\includegraphics[width=0.23\textwidth]{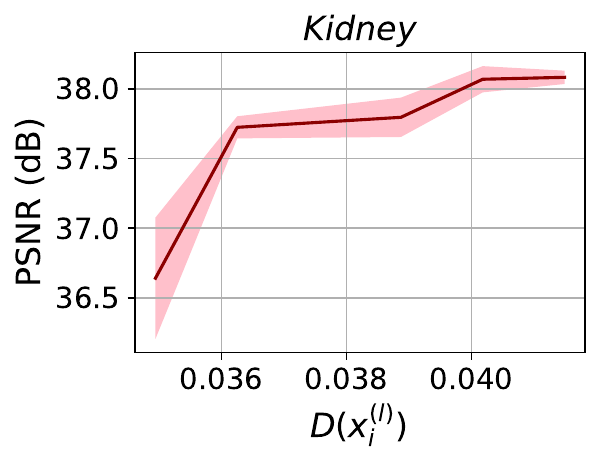}}
	\subfigure{
		\includegraphics[width=0.23\textwidth]{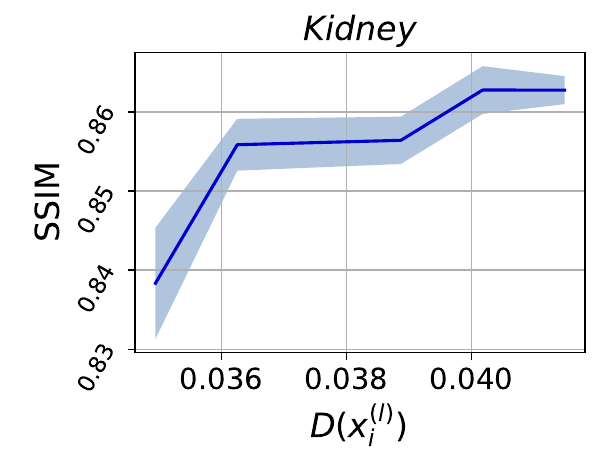}}
	\subfigure{
		\includegraphics[width=0.23\textwidth]{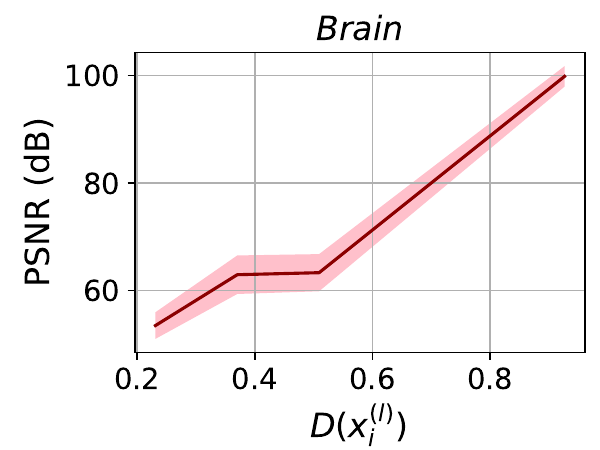}}
	\subfigure{
		\includegraphics[width=0.23\textwidth]{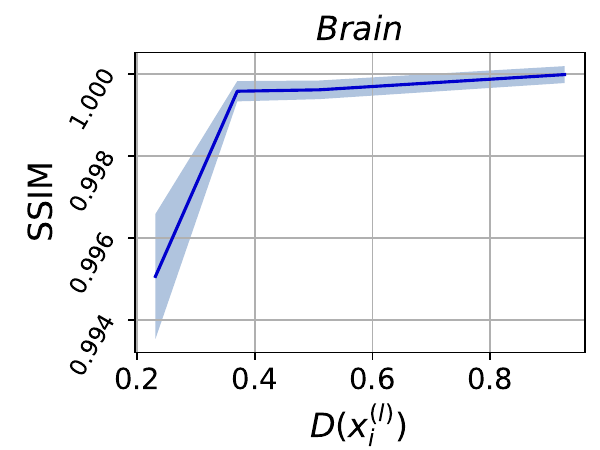}}
	\caption{Statistical graph of the relationship between reconstruction fidelity  and the degree of spectrum concentration on each dataset (denoted by $D(\boldsymbol{x}_{i}^{(l)})$). One can find a significant positive correlation between them.}
	\label{fig_sup: Statistical graph of the relationship between reconstruction fidelity  and the degree of spectrum concentration} 
\end{figure}

\newpage
\newpage
\newpage
\newpage
\newpage
\section*{Supplementary Tables}

\begin{table}[H]
\centering
    \caption{The ablation study of our model on each dataset. We used compression ratio $80\times$, data size $256\times256\times256$ voxels, iteration number 10,000 for medical data and compression ratio $100\times$, data size $16\times512\times512$ voxels, iteration number 10,000 for biological data.}
    \setlength{\tabcolsep}{0.25mm}{
    {\tabfontsize
    \begin{tabular}{l|ll|ll|ll|ll|ll|ll} \hline &
        \multicolumn{2}{l|}{\qquad\textit{Lung}} & \multicolumn{2}{l|}{\qquad\textit{Heart}} & 
        \multicolumn{2}{l|}{\qquad\textit{Kidney}} & 
        \multicolumn{2}{l|}{\qquad\textit{Brain}} & 
        \multicolumn{2}{l|}{\qquad\textit{Neurons}} & 
        \multicolumn{2}{l}{\qquad\textit{Vessels}} \\
        \cline{2-13}
        Method 
        & {\scriptsize PSNR} & ~~~{\scriptsize SSIM}
        & {\scriptsize PSNR} & ~~~{\scriptsize SSIM}
        & {\scriptsize PSNR} & ~~~{\scriptsize SSIM}
        & {\scriptsize PSNR} & ~~~{\scriptsize SSIM} 
        & {\scriptsize Acc.(200)} & ~{\scriptsize Acc.(500)}
        & {\scriptsize Acc.(200)} & ~{\scriptsize Acc.(500)}
        \\
    \hline
        \M~(ours) & \textbf{47.46} & \textbf{0.9820} & \textbf{59.49} & \textbf{0.9993} & \textbf{37.18} & \textbf{0.8464} & \textbf{53.02} & \textbf{0.9953} & \textbf{0.9944} & \textbf{0.9970} & \textbf{0.9560} &
        \textbf{0.9760} \\ 
    \hline
        EP & ~46.81 & ~0.9805 & ~58.84 & ~0.9992 & ~36.96 & ~0.8437 & ~52.06 & ~0.9950 & ~0.9940 & ~0.9967 & ~0.9471 &
        ~0.9716 \\ 
        NP & ~46.68 & ~0.9795 & ~55.95 & ~0.9989 & ~36.84 & ~0.8394 & ~51.58 & ~0.9948 & ~0.9880 & ~0.9922 & ~0.9460 &
        ~0.9747 \\ 
    \hline
        $fr=1.0$ & ~47.41 & ~0.9819 & ~59.44 & ~0.9993 & ~37.15 & ~0.8459 & ~52.97 & ~0.9952 & ~0.9922 & ~0.9955 & ~0.9151 &
        ~0.9521 \\ 
        $fr<1.0$ & ~47.05 & ~0.9807 & ~59.03 & ~0.9992 & ~36.91 & ~0.8414 & ~52.47 & ~0.9948 & ~0.9847 & ~0.9919 & ~0.8860 &
        ~0.9281 \\ 
    \hline
        By Size & ~47.41 & ~0.9818 & ~58.52 & ~0.9992 & ~37.13 & ~0.8474 & ~52.97 & ~0.9953 & ~0.9939 & ~0.9966 & ~0.9476 &
        ~0.9731 \\ 
        By $D$ & ~47.27 & ~0.9815 & ~57.97 & ~0.9992 & ~37.10 & ~0.8472 & ~52.95 & ~0.9954 & ~0.9942 & ~0.9968 & ~0.9548 &
        ~0.9753 \\ 
        Equal & ~47.04 & ~0.9809 & ~56.60 & ~0.9990 & ~37.01 & ~0.8467 & ~52.74 & ~0.9951 & ~0.9941 & ~0.9967 & ~0.9473 &
        ~0.9742 \\ 
    \hline
    \end{tabular}}
    }
    \label{Table_sup: 1}
\end{table}

\begin{table}[H]
\centering
\caption{The mean of compression time and decompression time for each method on 6 neurons data, which were cropped to the same size ($16\times256\times256$ voxels), under around 1024$\times$ compression ratio. Comparatively, INR based compressor (SCI and NeRV) is slower in compression stage but can decompress faster. Since biomedical data need to be compressed only once before saving or transmission, but decompressed frequently for processing and analysis. Therefore, SCI is advantageous in biomedical fields.}
\begin{tabular}{ccc}
\hline
Method &
  \begin{tabular}[c]{@{}c@{}}Compression\\ (seconds)\end{tabular} &
  \begin{tabular}[c]{@{}c@{}}Decompression\\ (seconds)\end{tabular} \\ \hline
SCI (ours) & 40.1 & 0.011 \\
NeRV & 98.4 & 0.012 \\
HEVC & 0.165 & 0.064 \\ \hline
\end{tabular}
\label{tab: speed}
\end{table}
\end{document}